\documentclass[12pt]{article}
\usepackage{amssymb}

  \advance\oddsidemargin  by -1in
  \advance\evensidemargin by -1in
  \advance\topmargin      by -0.5in
\input epsf.tex

\newcommand*{\epsfscale}[1]{\def\epsfsize##1##2{#1##1}}

\newcommand{\Z}{\mathbb{Z}}
\newcommand{\C}{\mathbb{C}}
\newcommand{\LL}{\mathbf{L}}
\newcommand{\U}{\mathbf{U}}
\newcommand{\D}{\mathbf{D}}

\newcommand{\Om}{\Omega}
\newcommand{\Lam}{\Lambda}
\newcommand{\Del}{\mbox{\large$\delta$}}
\newcommand{\Gam}{\mbox{\large$\gamma$}}
\newcommand{\Eps}{\mbox{\large$\varepsilon$}}
\newcommand{\Tau}{\mbox{\large$\tau$}}
\newcommand{\Sig}{\mbox{\large$\sigma$}}
\newcommand{\rE}{\mbox{\large$\mathrm e$}}
\newcommand{\rC}{\mbox{\large$\mathrm c$}}
\newcommand{\rS}{\mathrm{S}}
\newcommand{\rT}{\tilde{\mathrm{S}}}
\newcommand{\rR}{\mathrm{R}}
\newcommand{\rI}{\bar{\mathrm{R}}}

\newcommand{\del}{\delta} 

\newcommand{\Supp}{\mathop\mathrm{Supp}\nolimits}
\newcommand{\TOR}{\mathop\mathrm{TOR}\nolimits}

\newcommand*{\boldind}[1]{{\mbox{\scriptsize\boldmath$#1$}}}

\newcommand*{\newCOMMAND}[3]{\newcommand*{#1}[1]{\def\aA#2{#3}{\aA##1}}}

\newcommand*{\RD}[1]{D_\boldind{#1}}
\newcommand*{\RF}[1]{F^\boldind{#1}}
\newCOMMAND{\OM}{##1##2##3}{\Om^\boldind{##1}_\boldind{##2##3}}
\newCOMMAND{\LAM}{##1##2##3}{\Lam^\boldind{{##1}{##2}}_\boldind{##3}}
\newCOMMAND{\DEL}{##1##2}{\Del^\boldind{##1}_\boldind{##2}}
\newcommand*{\EPS}[1]{\Eps_\boldind{#1}}
\newcommand*{\TAU}[1]{\Tau_\boldind{#1}}
\newcommand*{\RE}[1]{\rE^\boldind{#1}}
\newcommand*{\RC}[1]{\rC^\boldind{#1}}
\newCOMMAND{\RS}{##1##2}{\rS^\boldind{##1}_\boldind{##2}}
\newCOMMAND{\RT}{##1##2}{\rT^\boldind{##1}_\boldind{##2}}
\newCOMMAND{\RR}{##1##2}{\rR^\boldind{##1##2}}
\newCOMMAND{\RI}{##1##2}{\rI^\boldind{##1##2}}
\newcommand*{\PSI}[1]{|\psi^\boldind{#1}\rangle}
\newCOMMAND{\cRR}{##1##2##3##4}{\calR^\boldind{##1##2}_\boldind{##3##4}}

\newcommand{\calA}{\mathcal{A}}
\newcommand{\calB}{\mathcal{B}}
\newcommand{\calC}{\mathcal{C}}
\newcommand{\calD}{\mathcal{D}}
\newcommand{\calE}{\mathcal{E}}
\newcommand{\calF}{\mathcal{F}}
\newcommand{\calG}{\mathcal{G}}
\newcommand{\calH}{\mathcal{H}}
\newcommand{\calI}{\mathcal{I}}
\newcommand{\calK}{\mathcal{K}}
\newcommand{\calL}{\mathcal{L}}
\newcommand{\calM}{\mathcal{M}}
\newcommand{\calN}{\mathcal{N}}
\newcommand{\calP}{\mathcal{P}}
\newcommand{\calR}{\mathcal{R}}
\newcommand{\calT}{\mathcal{T}}
\newcommand{\calU}{\mathcal{U}}

\title{Fault-tolerant quantum computation by anyons}

\author{ \textbf{A.~Yu.~Kitaev}\medskip\\
\textit{L.D.Landau Institute for Theoretical Physics,} \\
\textit{117940, Kosygina St.~2}\smallskip \\
e-mail:\quad kitaev\,@\,itp.ac.ru
\vspace{1cm} }

\begin{document}

\maketitle

\begin{abstract}
A two-dimensional quantum system with anyonic excitations can be considered
as a quantum computer. Unitary transformations can be performed by moving
the excitations around each other. Measurements can be performed by joining
excitations in pairs and observing the result of fusion. Such 
computation is fault-tolerant by its physical nature.
\end{abstract}

A quantum computer can provide fast solution for certain computational
problems (e.g. factoring and discrete logarithm~\cite{Shor1}) which require
exponential time on an ordinary computer. Physical realization of a quantum
computer is a big challenge for scientists. One important problem is
decoherence and systematic errors in unitary transformations which occur in
real quantum systems. From the purely theoretical point of view, this problem
has been solved due to Shor's discovery of fault-tolerant quantum
computation~\cite{Shor2}, with subsequent
improvements~\cite{KL,KLZ,Ah,Kit1}. An arbitrary quantum circuit
can be simulated using imperfect gates, provided these gates are close to the
ideal ones up to a constant precision $\delta$. Unfortunately, the threshold
value of $\delta$ is rather small\footnote{ Actually, the threshold is not
known. Estimates vary from $1/300$~\cite{Zalka} to $10^{-6}$~\cite{KLZ}.}; 
it is very difficult to achieve this precision.

Needless to say, \emph{classical} computation can be also performed 
fault-tolerantly. However, it is rarely done in practice because classical 
gates are reliable enough. Why is it possible? Let us try to 
understand the easiest thing --- why classical information can be stored 
reliably on a magnetic media. Magnetism arise from spins of individual atoms. 
Each spin is quite sensitive to thermal fluctuations. But the spins interact 
with each other and tend to be oriented in the same direction. If some spin 
flips to the opposite direction, the interaction forces it to flip back to the 
direction of other spins. This process is quite similar to the standard error 
correction procedure for the repetition code. We may say that errors are 
being corrected at the physical level. Can we propose something similar in 
the quantum case? Yes, but it is not so simple. First of all, we need a 
quantum code with local stabilizer operators. 

I start with a class of stabilizer quantum codes associated with lattices on
the torus and other 2-D surfaces~\cite{Kit1,Kit2}. Qubits live on the 
edges of the lattice whereas the stabilizer operators correspond to the 
vertices and the faces. These operators can be put together to make up a 
Hamiltonian with local interaction. (This is a kind of penalty function; 
violating each stabilizer condition costs energy). The ground state of this 
Hamiltonian coincides with the protected space of the code. It is $4^g$-fold 
degenerate, where $g$ is the genus of the surface. The degeneracy is 
persistent to local perturbation. Under small enough perturbation, the 
splitting of the ground state is estimated as $\exp(-aL)$, where $L$ is the 
smallest dimension of the lattice. This model may be considered as a quantum 
memory, where stability is attained at the physical level rather than by an 
explicit error correction procedure.

Excitations in this model are anyons, meaning that the global wavefunction 
acquires some phase factor when one excitation moves around the other. One 
can operate on the ground state space by creating an excitation pair, moving 
one of the excitations around the torus, and annihilating it with the other 
one. Unfortunately, such operations do not form a complete basis. It seems 
this problem can be removed in a more general model (or models) where the 
Hilbert space of a qubit have dimensionality $>2$. This model is related to 
Hopf algebras.

In the new model, we don't need torus to have degeneracy. An $n$-particle
excited state on the plane is already degenerate, unless the particles
(excitations) come close to each other. These particles are nonabelian anyons,
i.e. the degenerate state undergoes a nontrivial unitary transformation when
one particle moves around the other. Such motion (``braiding'') can be 
considered as fault-tolerant quantum computation. A measurement of the final 
state can be performed by joining the particles in pairs and observing the 
result of fusion.

Anyons have been studied extensively in the field-theoretic
context~\cite{Wil,DPR,BDP,BP,LP}. So, I hardly discover any new about their
algebraic properties. However, my approach differs in several respects:
\begin{itemize}
\item The model Hamiltonians are different.
\item We allow a generic (but weak enough) perturbation which removes 
\emph{any} symmetry of the Hamiltonian.\footnote{
 Some local symmetry still can be established by adding unphysical degrees of 
freedom, see Sec.~\ref{sec_matsymm}.}
\item The language of ribbon and local operators (see Sec.~\ref{sec_ribop}) 
provides unified description of anyonic excitations and long range 
entanglement in the ground state.
\end{itemize}

An attempt to use \emph{one-dimensional} anyons for quantum computation was
made by G.~Castagnoli and M.~Rasetti~\cite{CaRa}, but the question of 
fault-tolerance was not considered.

\section{Toric codes and the corresponding Hamiltonians} \label{sec_torcodes}

\begin{figure}[t]
\centerline{\epsfscale{0.6}\epsffile{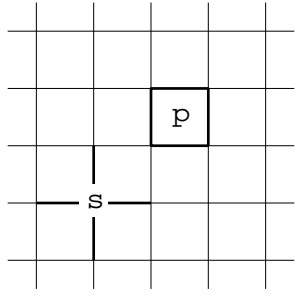}}
\caption{Square lattice on the torus}
\label{fig_sqlatt}
\end{figure}

Consider a $k\times k$ square lattice on the torus (see 
fig.~\ref{fig_sqlatt}). Let us attach a spin, or qubit, to each edge of 
the lattice. (Thus, there are $n=2k^2$ qubits). For each vertex $s$ and each 
face $p$, consider operators of the following form
\begin{equation} \label{stabop}
  A_s\ =\ \prod_{j\in\mbox{\scriptsize star}(s)}\sigma^x_j  \qquad\qquad  
  B_p\ =\ \prod_{j\in\mbox{\scriptsize boundary}(p)}\sigma^z_j
\end{equation} 
These operators commute with each other because $\mbox{star}(s)$ and 
$\mbox{boundary}(p)$ have either 0 or 2 common edges. The operators $A_s$ 
and $B_p$ are Hermitian and have eigenvalues $1$ and $-1$. 

Let $\calN$ be the Hilbert space of all $n=2k^2$ qubits. Define a 
\emph{protected subspace} $\calL\subseteq\calN$ as follows\footnote{
 We will show that this subspace is really protected from certain errors.
 Vectors of this subspaces are supposed to represent ``quantum 
 information'', like codewords of a classical code represent classical 
 information.}
\begin{equation} \label{infspace}
  \calL\ =\ \Bigl\{\, |\xi\rangle\in\calN\,:\,\  
  A_s|\xi\rangle=|\xi\rangle,\ B_p|\xi\rangle=|\xi\rangle\ \mbox{for all}\ s,p
  \,\Bigr\}
\end{equation}
This construction gives us a definition of a quantum code 
$\TOR(k)$, called a \emph{toric code}~\cite{Kit1,Kit2}. 
The operators $A_s$, $B_p$ are the \emph{stabilizer operators} of this code.

To find the dimensionality of the subspace $\calL$, we can observe that there 
are two relations between the stabilizer operators,\, $\prod_s A_s=1$\, and 
$\prod_p B_p=1$. So, there are $m=2k^2-2$ independent stabilizer operators. 
It follows from the general theory of additive quantum codes~\cite{Got,CRSS} 
that\, $\dim\calL=2^{n-m}=4$.

\begin{figure}[t]
\centerline{\epsfscale{0.6}\epsffile{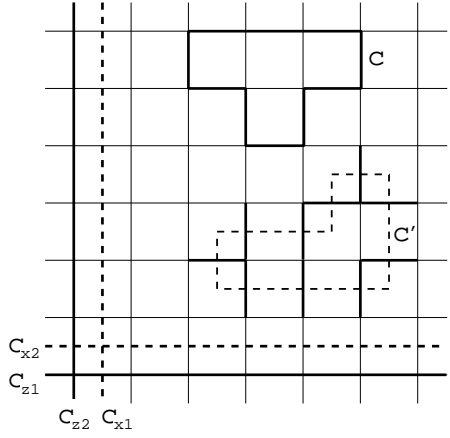}}
\caption{Loops on the lattice and the dual lattice}
\label{fig_loops}
\end{figure}

However, there is a more instructive way of computing $\dim\calL$. Let us find
the algebra $\LL(\calL)$ of all linear operators on the space $\calL$ --- this
will give us full information about this space. Let $\calF\subseteq\LL(\calN)$ 
be the algebra of operators generated by $A_s$, $B_p$. Clearly, 
$\LL(\calL)\cong\calG/\calI$,\, where $\calG\supseteq\calF$ is the algebra of 
all operators which commute with $A_s$, $B_p$,\, and $\calI\subset\calG$ is 
the ideal generated by $A_s\!-\!1$,\, $B_p\!-\!1$. The algebra $\calG$ is 
generated by operators of the form
\[
  Z\ =\ \prod_{j\in c}\sigma^z_j \qquad\qquad X\ =\ \prod_{j\in c'}\sigma^x_j
\]
where $c$ is a loop (closed path) on the lattice, whereas $c'$ is a cut, i.e. 
a loop on the dual lattice (see fig.~\ref{fig_loops}). If a loop (or a cut)
is contractible then the operator $Z$ is a product of $B_p$, hence 
$Z\equiv 1\pmod{\calI}$. Thus, only non-contractible loops or cuts are 
interesting. It follows that the algebra $\LL(\calL)$ is generated by 
4 operators $Z_1$, $Z_2$, $X_1$, $X_2$ corresponding  to the loops $c_{z1}$, 
$c_{z2}$ and the cuts $c_{x1}$, $c_{x2}$ (see fig.~\ref{fig_loops}). 
The operators $Z_1$, $Z_2$, $X_1$, $X_2$ have the same commutation relations 
as $\sigma^z_1$, $\sigma^z_2$, $\sigma^x_1$, $\sigma^x_2$. We see that each 
quantum state $|\xi\rangle\in\calL$ corresponds to a state of 2 qubits. Hence, 
the protected subspace $\calL$ is 4-dimensional. 

In a more abstract language, the algebra $\calF$ corresponds to $2$-boundaries 
and $0$-coboundaries (with coefficients from $\Z_2$), $\calG$ corresponds to 
$1$-cycles and $1$-cocycles, and $\LL(\calL)$ corresponds to $1$-homologies
and $1$-cohomologies.

There is also an explicit description of the protected subspace which may be 
not so useful but is easier to grasp. Let us choose basis vectors in the 
Hilbert space $\calN$ by assigning a label $z_j=0,1$ to each edge 
$j$.~\footnote{
 $0$ means ``spin up'', $1$ means ``spin down''. The Pauli operators 
 $\sigma_z$, $\sigma_x$ have the standard form in this basis.}
The constraints $B_p|\xi\rangle=|\xi\rangle$ say that the sum of the labels 
at the boundary of a face should be zero $(\bmod\ 2)$. More exactly, only such 
basis vectors contribute to a vector from the protected subspace. Such a 
basis vector is characterized by two topological numbers: the sums of $z_j$ 
along the loops $c_{z1}$ and $c_{z2}$. The constraints 
$A_p|\xi\rangle=|\xi\rangle$ say that all basis vectors with the same 
topological numbers enter $|\xi\rangle$ with equal coefficients. 
Thus, for each of the $4$ possible combinations of the topological numbers 
$v_1,v_2$, there is one vector from the protected subspace,
\begin{equation} \label{grstate}
  |\xi_{v_1,v_2}\rangle\ =\ 
  2^{-(k^2-1)/2}\, \sum_{z_1,\ldots,z_n}|z_1,\ldots,z_n\rangle\quad : \qquad
  \sum_{j\in c_{z1}}z_j=v_1,\quad \sum_{j\in c_{z2}}z_j=v_2
\end{equation}
Of course, one can also create linear combinations of these vectors.

Now we are to show that the code $\TOR(k)$ detects $k-1$ errors\footnote{
 In the theory of quantum codes, the word ``error'' is used in a somewhat 
 confusing manner. Here it means a single qubit error. In most other cases, 
 like in the formula below, it means a multiple error, i.e.\ an arbitrary 
 operator $E\in\LL(\calN)$.} 
(hence, it corrects $\left\lfloor\frac{k-1}{2}\right\rfloor$ errors). 
Consider a multiple error
\[
  E\ =\ \sigma(\alpha_1,\ldots,\alpha_n;\,\beta_1,\ldots,\beta_n)\ =\,\ 
  \prod_j(\sigma^x_j)^{\alpha_j}\ \prod_j(\sigma^z_j)^{\beta_j}
  \qquad (\alpha_j,\beta_j=0,1)
\]
This error can not be detected  by syndrome measurement (i.e. by measuring 
the eigenvalues of all $A_s$, $B_j$) if and only if $E\in\calG$. However, if 
$E\in\calF$ then $E|\xi\rangle=|\xi\rangle$ for every $|\xi\rangle\in\calL$. 
Such an error is not an error at all --- it does not affect the protected 
subspace. The bad case is when $E\in\calG$ but $E\not\in\calF$. Hence, the 
support of $E$ should contain a non-contractible loop or cut. It is only 
possible if $|\Supp(E)|\ge k$. (Here $\Supp(E)$ is the set of $j$ 
for which $\alpha_j\not=0$ or $\beta_j\not=0$).

One may say that the toric codes have quite poor parameters. Well, they are 
not ``good'' codes in the sense of~\cite{CaSh}. However, the code 
$\TOR(k)$ corrects \emph{almost any} multiple error of size $O(k^2)$. (The 
constant factor in $O(\ldots)$ is related to the percolation problem). So, 
these codes work if the error rate is constant but smaller than some 
threshold value. The nicest property of the codes $\TOR(k)$ is that they are 
\emph{local check codes}. Namely,
\begin{enumerate}
\item Each stabilizer operators involves bounded number of qubits 
(at most $4$).
\item Each qubit is involved in a bounded number of stabilizer operators
(at most $4$).
\item There is no limit for the number of errors that can be corrected.
\end{enumerate}
Also, at a constant error rate, the unrecoverable error probability goes to 
zero as $\exp(-ak)$.

It has been already mentioned that error detection involves syndrome 
measurement. To correct the error, one needs to find its characteristic vector 
$(\alpha_1,\ldots,\alpha_n;\,\beta_1,\ldots,\beta_n)$ out of the syndrome.
This is the usual error correction scheme. A new suggestion is to perform 
error correction at the physical level. Consider the Hamiltonian
\begin{equation} \label{Ham}
  H_0\ =\ -\sum_s A_s \,-\, \sum_p B_p
\end{equation}
Diagonalizing this Hamiltonian is an easy job because the operators $A_s$, 
$B_p$ commute. In particular, the ground state coincides with the 
protected subspace of the code $\TOR(k)$; it is 4-fold degenerate.
All excited states are separated by an energy gap $\Delta E\ge 2$, because 
the difference between the eigenvalues of $A_s$ or $B_p$ equals $2$. This 
Hamiltonian is more or less realistic because in involves only \emph{local}
interactions. We can expect that ``errors'', i.e.\ noise-induced excitations 
will be removed automatically by some relaxation processes. Of course, this 
requires cooling, i.e.\ some coupling to a thermal bath with low temperature 
(in addition to the Hamiltonian~(\ref{Ham})).

Now let us see whether this model is stable to perturbation. (If not, there is
no practical use of it). For example, consider a perturbation of the form
\[
  V\ =\ -\vec{h}\sum_j\vec{\sigma}_j\ -\ 
  \sum_{j<p}J_{jp}(\vec{\sigma}_j,\vec{\sigma}_p)
\]
It is important that the perturbation is local, i.e.\ each term of it contains
a small number of $\sigma$ (at most 2). Let us estimate the energy splitting 
between two orthogonal ground states of the original Hamiltonian, 
$|\xi\rangle\in\calL$ and $|\eta\rangle\in\calL$. We can use the usual 
perturbation theory because the energy spectrum has a gap. In the $m$-th order 
of the perturbation theory, the splitting is proportional to\, 
$\langle\xi|V^m|\eta\rangle$\, or\, 
$\langle\xi|V^m|\xi\rangle-\langle\eta|V^m|\eta\rangle$. However, both 
quantities are zero unless $V^m$ contains a product of $\sigma^z_j$ or 
$\sigma^x_j$ along a non-contractible loop or cut. Hence, the splitting 
appears only in the $\lceil k/2\rceil$-th or higher orders. As far as all 
things (like the number of the relevant terms in $V^m$) scale correctly to 
the thermodynamic limit, the splitting vanishes as $\exp(-ak)$. A simple 
physical interpretation of this result is given in the next section.
(Of course, the perturbation should be small enough, or else a phase 
transition may occur).

Note that our construction is not restricted to square lattices. We can 
consider an arbitrary irregular lattice, like in fig.~\ref{fig_lattice}. 
Moreover, such a lattice can be drawn on an arbitrary 2-D surface. On a
compact orientable surface of genus $g$, the ground state is $4^g$-fold 
degenerate. In this case, the splitting of the ground state is estimated as 
$\exp(-aL)$, where $L$ is the smallest dimension of the lattice. We see that 
the ground state degeneracy depends on the surface topology, so we deal with 
\emph{topological quantum order}. On the other hand, there is a finite energy 
gap between the ground state and excited states, so all spatial correlation 
functions decay exponentially. This looks like a paradox --- 
how do parts of a macroscopic system know about the topology if all 
correlations are already lost at small scales? The answer is that there 
is long-range entanglement~\footnote{
 Entanglement is a special, purely quantum form of correlation.} 
which can not be expressed by simple correlation functions like 
$\langle\sigma^a_j\sigma^b_l\rangle$. This entanglement reveals itself in the 
excitation properties we are going to discuss.

\section{Abelian anyons} \label{sec_abanyons}

\begin{figure}[t]
\centerline{\epsfscale{0.6}\epsffile{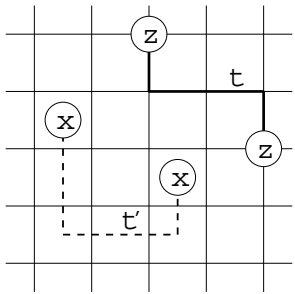}}
\caption{Strings and particles}
\label{fig_strings}
\end{figure}

Let us classify low-energy excitations of the Hamiltonian~(\ref{Ham}). An 
eigenvector of this Hamiltonian is an eigenvector of all the operators 
$A_s$, $B_p$. An \emph{elementary excitation}, or \emph{particle} occurs  if 
only one of the constraints\, $A_s|\xi\rangle=|\xi\rangle$,\,\, 
$B_p\xi\rangle=|\xi\rangle$\, is violated. Because of the relations 
$\prod_s A_s=1$ and $\prod_p B_p=1$,\, it is impossible to create a single 
particle. However, it is possible to create a two-particle state of the form
$|\psi^z(t)\rangle=S^z(t)|\xi\rangle$ 
or $|\psi^x(t')\rangle=S^x(t')|\xi\rangle$, 
where $|\xi\rangle$ is an arbitrary ground state, and
\begin{equation} \label{stringop}
  S^z(t)\ =\ \prod_{j\in t}\sigma^z_j \qquad\qquad\quad 
  S^x(t')\ =\ \prod_{j\in t'}\sigma^x_j
\end{equation}
(see fig.~\ref{fig_strings}). In the first case, two particles are created at 
the endpoints of the ``string'' (non-closed path) $t$. Such particles live on 
the vertices of the lattice. We will call them $z$-type particles, or 
``electric charges''. Correspondingly, $x$-type particles, or 
``magnetic vortices'' live on the faces. The operators $S^z(t)$, 
$S^x(t')$ are called \emph{string operators}. Their characteristic property is 
as follows:\, they commute with every $A_s$ and $B_p$, except for few ones 
(namely, 2) corresponding to the endpoints of the string. Note that the state 
$|\psi^z(t)\rangle=S^z(t)|\xi\rangle$ depends only on the homotopy class of 
the path $t$ while the operator $S^z(t)$ depends on $t$ itself.

Any configuration of an even number of $z$-type particles and an even number 
of $x$-type particles is allowed. We can connect them by strings in an 
arbitrary way. Each particle configuration defines a $4^g$-dimensional 
subspace in the global Hilbert space $\calN$. This subspace is independent of 
the strings but a particular vector 
$S^{a_1}(t_1)\cdots S^{a_m}(t_m)|\xi\rangle$ depends on $t_1,\ldots,t_m$. 
If we draw these strings in a topologically different way, we get another 
vector in the same $4^g$-dimensional subspace. Thus, the strings are 
unphysical but we can not get rid of them in our formalism. 

Let us see what happens if these particles move around the torus (or other
surface). Moving a $z$-type particle along the path $c_{z1}$ or $c_{z2}$ 
(see fig.~\ref{fig_loops}) is equivalent to applying the operator $Z_1$ or 
$Z_2$. Thus, we can operate on the ground state space by creating a particle
pair, moving one of the particles around the torus, and annihilating it with 
the other one. Thus we can realize some quantum gates. Unfortunately, 
too simple ones --- we can only apply the operators $\sigma_z$ and $\sigma_x$ 
to each of the $2$ (or $2g$) qubits encoded in the ground state.

Now we can give the promised physical interpretation of the ground state 
splitting. In the presence of perturbation, the two-particle state 
$|\psi^z(t)\rangle$ is not an eigenstate any more. More exactly, both 
particles will propagate rather than stay at the same positions. The 
propagation process is described by the Schr\"odinger equation with some 
effective mass $m_z$. ($x$-type particles have another mass $m_x$). In the 
non-perturbed model, $m_z=m_x=\infty$. There are no real particles in the 
ground state, but they can be created and annihilate virtually. A virtual 
particle can tunnel around the torus before annihilating with the other one. 
Such processes contribute terms 
$b_{z1}Z_1$, $b_{z2}Z_2$, $b_{x1}X_1$, $b_{x2}X_2$ to the ground state 
effective Hamiltonian. Here $b_{\alpha i}\sim\exp(-a_\alpha L_i)$ is the 
tunneling amplitude whereas $a_\alpha\sim\sqrt{2m\Delta\!E}$ is the imaginary 
wave vector of the tunneling particle.

\begin{figure}[t]
\centerline{\epsfscale{0.6}\epsffile{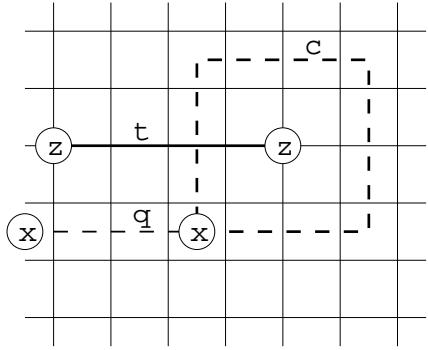}}
\caption{An $x$-type particle moving around a $z$-type particle}
\label{fig_movearou}
\end{figure}

Next question: what happens if we move particles around each other? (For this, 
we don't need a torus; we can work on the plane). For example, let us move 
an $x$-type particle around a $z$-type particle (see fig.~\ref{fig_movearou}). 
Then 
\[
  |\Psi_{\mbox{\scriptsize initial}}\rangle\, =\, S^z(t)\,|\psi^x(q)\rangle
  \ ,\qquad\quad
  |\Psi_{\mbox{\scriptsize final}}\rangle\, =\, 
  S^x(c)\,S^z(t)\,|\psi^x(q)\rangle\ =\ 
  -|\Psi_{\mbox{\scriptsize initial}}\rangle
\]
because $S^x(c)$ and $S^z(t)$ anti-commute, and 
$S^x(c)|\psi^x(q)\rangle=|\psi^x(q)\rangle$.
We see that the global wave function ($=$ the state of the entire system) 
acquires the phase factor $-1$. It is quite unlike usual particles, bosons 
and fermions, which do not change their phase in such a process. Particles 
with this unusual property are called \emph{abelian anyons}. More generally, 
abelian anyons are particles which realize nontrivial one-dimensional 
representations of (colored) braid groups. In our case, the phase change 
can be also interpreted as an Aharonov-Bohm effect. It does not occur if both 
particles are of the same type.

Note that abelian anyons exist \emph{in real solid state systems}, namely, 
they are intrinsicly related to the fractional quantum Hall effect~\cite{ASW}. 
However, these anyons have different braiding properties. In the fractional 
quantum Hall system with filling factor $p/q$, there is only one basic type of 
anyonic particles with (real) electric charge $1/q$. (Other particles are 
thought to be composed from these ones). When one particle moves around the 
other, the wave function acquires a phase factor $\exp(2\pi i/q)$.

Clearly, the existence of anyons and the ground state degeneracy have the same 
nature. They both are manifestations of a topological quantum order, a hidden 
long-range order that can not be described by any local order parameter. (The 
existence of a local order parameter contradicts the nature of a quantum 
code --- if the ground state is accessible to local measurements then it is 
not protected from local errors). It seems that the anyons are more 
fundamental and can be used as a universal probe for this hidden order. 
Indeed, \emph{the ground state degeneracy on the torus follows from the 
existence of anyons}~\cite{Ein}. Here is the original Einarsson's proof 
applied to our two types of particles.

We derived the ground state degeneracy from the commutation relations between 
the operators $Z_1$, $Z_2$, $X_1$, $X_2$. These operators can be realized by 
moving particles along the loops $c_{z1}$, $c_{z2}$, $c_{x1}$, $c_{x2}$. These 
loops only exist on the torus, not on the plane. Consider, however, the 
process in which an $x$-type particle and a $z$-type particle go around the 
torus and then trace their paths backward. This corresponds to the operator 
$W=X_1^{-1}Z_1^{-1}X_1Z_1$ which can be realized on the plane. Indeed, we can 
deform particles' trajectories  so that one particle stays at rest while the 
other going around it. Due to the anyonic nature of the particles, $W=-1$. 
We see that $X_1$ and $Z_1$ anti-commute.

\begin{figure}[t]
\centerline{\epsfscale{0.6}\epsffile{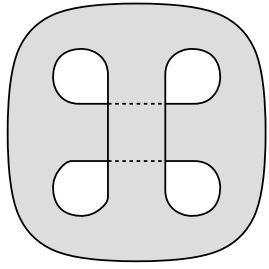}}
\caption{A fly-over crossing geometry for a 2-D electron layer}
\label{fig_crossing}
\end{figure}

The above argument is also applicable to the fractional quantum Hall
anyons~\cite{Ein}. The ground state on the torus is $q$-fold degenerate, up to
the precision $\sim\exp(-L/l_0)$, where $l_0$ is the magnetic length. This
result does not rely on the magnetic translational symmetry or any other
symmetry. Rather, it relies on the existence of the energy gap in the spectrum
(otherwise the degeneracy would be unstable to perturbation). Note that holes
($=$punctures) in the torus do not remove the degeneracy unless they break the
nontrivial loops $c_{x1}$, $c_{x2}$, $c_{z1}$, $c_{z2}$.
The fly-over crossing geometry (see fig.~\ref{fig_crossing}) is 
topologically equivalent to a torus with 2 holes, but it is almost flat. 
In principle, such structure can be manufactured~\footnote{
  It is not easy. How will the two layers (the two crossing 
  ``roads'', one above the other) join in a single crystal layout?},
cooled down and placed into a perpendicular magnetic field. This will be a 
sort of quantum memory --- it will store a quantum state forever, provided all 
anyonic excitation are frozen out or localized. Unfortunately, I do not know 
any way this quantum information can get in or out. Too few things can be done 
by moving abelian anyons. All other imaginable ways of accessing the ground 
state are uncontrollable.

\section{Materialized symmetry: is that a miracle?} \label{sec_matsymm}

Anyons have been studied extensively in the gauge field theory
context~\cite{Wil,DPR,BDP,LP}. However, we start with quite different
assumptions about the Hamiltonian. A gauge theory implies a gauge symmetry
which can not be removed by external perturbation. To the contrary, our model
is stable to \emph{arbitrary} local perturbations. It is useful to give a
field-theoretic interpretation of this model. The edge labels $z_j$
(measurable by $\sigma^z_j$) correspond to a $\Z_2$ vector potential, whereas
$\sigma^x_j$ corresponds to the electric field. The operators $A_s$ are local
gauge transformations whereas $B_p$ is the magnetic field on the face $p$.
The constraints $A_s|\xi\rangle=|\xi\rangle$ mean that the state $|\xi\rangle$
is gauge-invariant. Violating the gauge invariance is energetically
unfavorable but not forbidden. The Hamiltonian (which includes $H_0$ and some
perturbation $V$) need not obey the gauge symmetry. The constraints
$B_p|\xi\rangle=|\xi\rangle$ mean that the gauge field corresponds to a flat
connection. These constraints are not strict either.

Despite the absence of symmetry in the Hamiltonian $H=H_0+V$, 
our system exhibits two conservation laws: electric charge and magnetic charge 
(i.e.\ the number of vortices) are both conserved \emph{modulo} 2. 
In the usual electrodynamics, conservation of electric charge is related 
to the local ($=$gauge) $\U(1)$ symmetry. In our case, it should be a local 
$\Z_2$ symmetry for electric charges and another $\Z_2$ symmetry for magnetic 
vortices. So, our system exhibits a \emph{dynamically created} $\Z_2\times\Z_2$ 
symmetry which appears only at large distances where individual excitations 
are well-defined. 

Probably, the reader is not satisfied with this interpretation. Really, it 
creates a new puzzle rather than solve an old one. What is this mysterious 
symmetry? How do symmetry operators look like at the microscopic level? 
The answer sounds as nonsense but it is true. This symmetry (as well 
as any other local symmetry) can be found in \emph{any} Hamiltonian if we 
introduce some unphysical degrees of freedom. So, the symmetry is not actually 
being created. Rather, an artificial symmetry becomes a real one.

The new degrees of freedom are spin variables $v_s,w_p=0,1$ for each vertex
$s$ and each face $p$. The vertex spins will stay in the state
$2^{-1/2}\Bigl(|0\rangle+|1\rangle\Bigr)$ whereas the face spins will stay in
the state $|0\rangle$. So, all the extra spins together stay in a unique
quantum state $|\zeta\rangle$. Obviously,
$\sigma^x_s|\zeta\rangle=|\zeta\rangle$ and
$\sigma^z_p|\zeta\rangle=|\zeta\rangle$, for every vertex $s$ and every face
$p$. From the mathematical point of view, we have simply defined an embedding
of the space $\calN$ into a larger Hilbert space $\calT$ of all the spins,\,
$|\psi\rangle\mapsto|\psi\rangle\otimes|\zeta\rangle$.  So we may write
$\calN\subseteq\calT$. We will call $\calN$ the \emph{physical space} (or
subspace), $\calT$ the \emph{extended space}. Physical states (i.e.\ vectors
$|\psi\rangle\in\calN$) are characterized by the equations
\[
  \sigma^x_s|\psi\rangle\,=\,|\psi\rangle\ , \quad\ 
  \sigma^z_p|\psi\rangle\,=\,|\psi\rangle\ \qquad\quad
\]
for every vertex $s$ and face $p$.

Now let us apply a certain unitary transformation $U$ on the extended space 
$\calT$. This transformation is just a change of the spin variables, namely
\begin{equation}
  v_s\ \mapsto\ v_s\ , 
  \qquad\ 
  z_j\ \mapsto\ z_j\, + \sum_{s=\mbox{\scriptsize endpoint}(j)} v_s\ ,
  \qquad\ 
  w_p\ \mapsto\ w_p\, + \sum_{j\in\mbox{\scriptsize boundary}(p)} z_j
\end{equation}
(all sums are taken \emph{modulo} 2). The physical subspace becomes 
$\calN'=U\calN$. Vectors $|\psi\rangle\in\calN'$ are invariant under the 
following symmetry operators
\begin{equation}\label{symop}
  P_s\,=\, U\sigma^x_s U^\dagger\, =\ \sigma^x_s\, A_s \qquad\qquad 
  Q_p\,=\, U\sigma^z_p U^\dagger\, =\ \sigma^z_p\, B_p
\end{equation}
The transformed Hamiltonian $H'=UHU^\dagger$ commutes with these operators.
It is defined up to the equivalence\, $P_s\equiv 1$,\,\, $Q_p\equiv 1$.
In particular, 
\begin{equation} \label{Ham_gauge}
  H_0'\ =\ U\,H_0\,U^\dagger\ =\ H_0'\ \equiv\ 
  - \sum_{s}\sigma^x_s\, -  \sum_{p}\sigma^z_p
\end{equation}

In the field theory language, the vertex variables $v_s$ (or the operators 
$\sigma^z_s$) are a Higgs field. The operators $P_s$ are local gauge 
transformations. Thus, an arbitrary Hamiltonian can be written in a 
gauge-invariant form if we introduce additional Higgs fields. Of course, 
it is a very simple observation. The real problem is to understand how the 
artificial gauge symmetry ``materialize'', i.e.\ give rise to a physical 
conservation law. 

Electric charge at a vertex $s$ is given by the operator $\sigma^x_s$. The 
total electric charge on a compact surface is zero\footnote{
 Strictly speaking, the electric charge is not a numeric quantity; rather, it 
 is an irreducible representation of the group $\Z_2$. ``Zero'' refers to the 
 identity representation.}
because $\prod_{s}\sigma^x_s\equiv 1$. This is not a physically meaningful 
statement as it is. It is only meaningful if there are discrete charged 
particles. Then the charge is also conserved locally, in every scattering or 
fusion process. It is difficult to formulate this property in a mathematical 
language, but, hopefully, it is possible. (The problem is that particles are 
generally smeared and can propagate. Physically, particles are well-defined 
if they are stable and have finite energy gap). Alternatively, one can use 
various local and nonlocal order parameters to distinguish 
between phases with an unbroken symmetry, broken symmetry or confinement. 

The artificial gauge symmetry materialize for the 
Hamiltonian~(\ref{Ham_gauge}) but this is not the case for every Hamiltonian.
Let us try to describe possible symmetry breaking mechanisms in terms of local 
order parameters. If the gauge symmetry is broken then there is a nonvanishing 
vacuum average of the Higgs field, $\phi(s)=\langle\sigma^z_s\rangle\not=0$. 
Electric charge is not conserved any more. In other words, there is a Bose 
condensate of charged particles. Although the second $\Z_2$ symmetry is 
formally unbroken, free magnetic vortices do not exist. More exactly, magnetic 
vortices are confined. (The duality between symmetry breaking and confinement 
is well known~\cite{Hooft}). It is also possible that the second 
symmetry is broken, then electric charges are confined. From the physical 
point of view, these two possibilities are equivalent: 
there is no conservation law in the system.\footnote{
 The two possibilities only differ if an already materialized symmetry breaks 
 down at much large distances (lower energies).}

An interesting question is whether magnetic vortices can be confined without
the gauge symmetry being broken. Apparently, the answer is ``no''. The
consequence is significant: electric charges and magnetic vortices can not
exist without each other. It seems that materialized symmetry needs better
understanding; as presented here, it looks more like a miracle.

\section{The model based on a group algebra}

 From now on, we are constructing and studying nonabelian anyons which will 
allow universal quantum computation.

Let $G$ be a finite (generally, nonabelian) group. Denote by $\calH=\C[G]$ the 
corresponding group algebra, i.e.\ the space of formal linear combinations of 
group elements with complex coefficients. We can consider $\calH$ as a Hilbert 
space with a standard orthonormal basis $\Bigl\{|g\rangle:\, g\in G\Bigr\}$.
The dimensionality of this space is $N=|G|$. We will work with ``spins''
(or ``qubits'')  taking values in this space.\footnote{
 In the field theory language, the value of a spin can be interpreted as a 
 $G$ gauge field. However, we do \emph{not} perform symmetrization over gauge
 transformations.}
\emph{Remark}: This model can be generalized. One can take for $\calH$ any 
finite-dimensional Hopf algebra equipped with a Hermitian scalar product with 
certain properties. However, I do not want to make things too complicated.

To describe the model, we need to define 4 types of linear operators, 
$L_+^g$, $L_-^g$, $T_+^h$, $T_-^h$ acting on the space $\calH$. 
Within each type, they are indexed by group elements, $g\in G$ or $h\in G$. 
They act as follows
\begin{equation}
  \begin{array}{rclcrcl}
    L_+^g|z\rangle &=& |gz\rangle &\qquad\qquad&
    T_+^h|z\rangle &=& \del_{h,z}\,|z\rangle      \medskip\\
    L_-^g|z\rangle &=& |zg^{-1}\rangle &\qquad\qquad&
    T_-^h|z\rangle &=& \del_{h^{-1}\!,\,z}\,|z\rangle
  \end{array}
\end{equation}
(In the Hopf algebra context, the operators $L_+^g$, $L_-^g$, $T_+^h$, $T_-^h$ 
correspond to the left and right multiplications and left and right 
comultiplication, respectively). These operators  satisfy the following 
commutation relations
\begin{equation} \label{commLT}
  \begin{array}{rclcrcl}
    L_+^g\, T_+^h &=& T_+^{gh}\, L_+^g &\qquad\qquad&
    L_+^g\, T_-^h &=& T_-^{hg^{-1}} L_+^g     \medskip\\
    L_-^g\, T_+^h &=& T_+^{hg^{-1}} L_-^g &\qquad\qquad&
    L_-^g\, T_-^h &=& T_-^{gh}\, L_-^g
  \end{array}
\end{equation}

\begin{figure}[t]
\centerline{\epsfscale{0.6}\epsffile{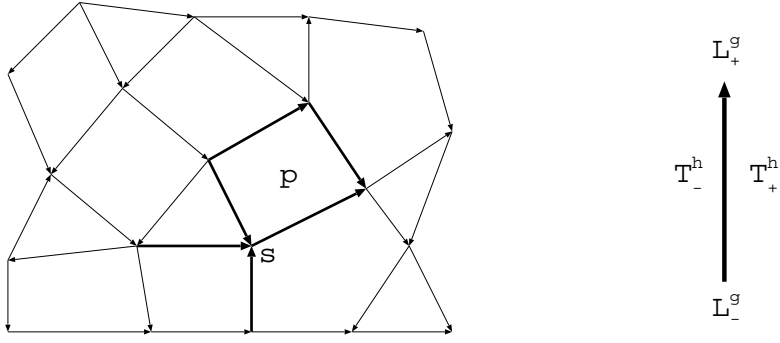}}
\caption{Generic lattice and the orientation rules for the operators 
         $L^g_\pm$ and $T^h_\pm$}
\label{fig_lattice}
\end{figure}

Now consider an arbitrary lattice on an arbitrary orientable 2-D surface, see
fig.~\ref{fig_lattice}. (We will mostly work with a plane or a sphere, not
higher genus surfaces). Corresponding to each edge is a spin which takes
values in the space $\calH$. Arrows in fig.~\ref{fig_lattice} mean that we
choose some orientation for each edge of the lattice. (Changing the direction
of a particular arrow will be equivalent to the basis change
$|z\rangle\mapsto|z^{-1}\rangle$ for the corresponding qubit). Let $j$ be an
edge of the lattice, $s$ one of its endpoints. Define an operator
$L^g(j,s)=L_\pm^g(j)$ as follows. If $s$ is the origin of the arrow $j$ then
$L^g(j,s)$ is $L_-^g(j)$ (i.e.\ $L_-^g$ acting on the $j$-th spin), otherwise
it is $L_+^g(j)$.  This rule is represented by the diagram at the right side
of fig.~\ref{fig_lattice}. Similarly, if $p$ is the left (the right) ajacent
face of the edge $j$ then $T^h(j,p)$ is $T_-^h$ (resp.\ $T_+^h$) acting on the
$j$-th spin.

Using these notations, we can define local gauge transformations and magnetic 
charge operators corresponding to a vertex $s$ and an adjacent face $p$ 
(see fig.~\ref{fig_lattice}). Put
\begin{equation} \label{locop}
  \begin{array}{rcl}
    A_g(s,p)&=& A_g(s)\ =\ 
    \prod\limits_{j\in\mbox{\scriptsize star}(s)}\,L^g(j,s) \medskip \\ 
    B_h(s,p) &=& 
    \sum\limits_{h_{1}\cdots h_{k}=h}\, \prod\limits_{m=1}^{k}\,T^{h_m}(j_m,p)
  \end{array}
\end{equation}
where $j_1,\ldots,j_k$ are the boundary edges of $p$ listed in the 
counterclockwise order, starting from, and ending at, the vertex $s$. 
(The sum is taken over all combinations of $h_1,\ldots,h_k\in G$, such that 
$h_1\cdots h_k=h$. Order is important here!). Although $A_g(s,p)$ does not 
depend on $p$, we retain this parameter to emphasize the duality between
$A_g(s,p)$ and $B_h(s,p)$.~\footnote{
  In the Hopf algebra setting, $A_g(s,p)$ does depend on $p$.}
These operators generate an algebra $\calD=\D(G)$, Drinfield's quantum 
double~\cite{Drin} of the group algebra $\C[G]$. It will play a very 
important role below. Now we only need two symmetric combinations of 
$A_g(s,p)$ and $B_h(s,p)$, namely
\begin{equation}
  A(s)\ =\ N^{-1} \sum_{g\in G} A_g(s,p) \qquad\qquad
  B(p)\ =\ B_1(s,p)
\end{equation}
where $N=|G|$. Both $A(s)$ and $B(p)$ are projection operators. ($A(s)$ 
projects out the states which are gauge invariant at $s$, whereas $B(p)$ 
projects out the states with vanishing magnetic charge at $p$).
The operators $A(s)$ and $B(p)$ commute with each other.\footnote{
 This is not obvious. Use the commutation relations~(\ref{commLT}) to verify 
 this statement.} 
Also $A(s)$ commutes with $A(s')$, and $B(p)$ commutes with $B(p')$ for 
different vertices and faces. In the case $G=\Z_2$, these operators are
almost the same as the operators~(\ref{stabop}),\, namely\,\, 
$A(s)=\frac{1}{2}(A_s+1)$,\,\,\, $B(p)=\frac{1}{2}(B_p+1)$.~\footnote{
 Here $A_s$ and $B_p$ are the notations from Sec.~\ref{sec_torcodes}; 
 we will not use them any more.}

At this point, we have only defined the global Hilbert space $\calN$ 
(the tensor product of many copies of $\calH$) and some operators on it. 
Now let us define the Hamiltonian.
\begin{equation} \label{Ham1}
  H_0\ =\ \sum_s (1-A(s)) \,+\, \sum_p (1-B(p))
\end{equation}
It is quite similar to the Hamiltonian~(\ref{Ham}). As in that case, the
space of ground states is given by the formula
\begin{equation} \label{grspace}
  \calL\ =\ \Bigl\{\, |\xi\rangle\in\calN\,:\,\  
  A(s)|\xi\rangle=|\xi\rangle,\ B(p)|\xi\rangle=|\xi\rangle\ 
  \mbox{for all}\ s,p  \,\Bigr\}
\end{equation}
The corresponding energy is $0$; all excited states have energies $\ge 1$.

It is easy to work out an explicit representation of ground states similar to
eq.~(\ref{grstate}). The ground states correspond 1-to-1 to flat
$G$-connections, defined up to conjugation, or superpositions of those. So,
the ground state on a sphere is not degenerate. However, particles
(excitations) have quite interesting properties even on the sphere or on the
plane. (We treat the plane as an infinitely large sphere). The reader probably
wants to know the answer first, and then follow formal calculations. So, I
give a brief abstract description of these particles. It is a mixture of
general arguments and details which require verification.

The particles live on vertices or faces, or both; in general, one particle
occupies a vertex and an adjacent face same time. A combination of a vertex
and an adjacent face will be called a \emph{site}. Sites are represented by
dotted lines in fig.~\ref{fig_ribbon}. (The dashed lines are edges of the dual
lattice).

Consider $n$ particles on the sphere pinned to particular sites
$x_1,\ldots,x_n$ at large distances from each other. The space
$\calL[n]=\calL(x_1,\ldots,x_n)$ of $n$-particle states has dimensionality
$N^{2(n-1)}$, including the ground state.\footnote{ The absence of particle at
a given site is regarded as a particle of special type.}  Not all these states
have the same energy. Even more splitting occurs under perturbation, but some
degeneracy still survive. Of course, we assume that the perturbation is local,
i.e.\ it can be represented by a sum of operators each of which acts only on
few spins. To find the residual degeneracy, we will study the action of such
\emph{local operators} on the space $\calL[n]$.  Local operators generate a
subalgebra $\calP[n]\subseteq\LL(\calL[n])$.  Elements of its center,
$\calC[n]$, are conserved classical quantities; they can be measured once and
never change. (More exactly, they can not be changed by local operators). As
these classical variables are locally measurable, we interpret them as
particle's types.  It turns out that the types correspond 1-to-1 to
irreducible representations of the algebra $\calD$, the quantum double. Thus,
each particle can belong to one of these types. The space $\calL[n]$ and the
algebra $\calP[n]$ split accordingly:
\begin{equation} 
  \calL[n]\ =\ \bigoplus_{d_1,\ldots,d_n} \calL_{d_1,\ldots,d_n}
  \qquad\quad
  \calP[n]\ =\ \bigoplus_{d_1,\ldots,d_n} \calP_{d_1,\ldots,d_n}
  \qquad\quad \Bigl(\, 
     \calP_{d_1,\ldots,d_n} \subseteq \LL(\calL_{d_1,\ldots,d_n})
  \,\Bigr)
\end{equation} 
where $d_m$ is the type of the $m$-th particle. The ``classical'' subalgebra  
$\calC[n]$ is generated by the projectors onto 
$\calL_{d_1,\ldots,d_n}$.

But this is not the whole story. The subspace $\calL_{d_1,\ldots,d_n}$ 
splits under local perturbations from $\calP_{d_1,\ldots,d_n}$. 
By a general mathematical argument,\footnote{
  $\calP_{d_1,\ldots,d_n}$ is a subalgebra of 
  $\LL(\calL_{d_1,\ldots,d_n})$ with a trivial center, 
  closed under Hermitian conjugation.}
this algebra can be characterized as follows
\begin{equation}
  \calL_{d_1,\ldots,d_n}\ =\ 
  \calK_{d_1,\ldots,d_n} \otimes \calM_{d_1,\ldots,d_n}
  \qquad\qquad
  \calP_{d_1,\ldots,d_n}\ =\ \LL(\calK_{d_1,\ldots,d_n})
\end{equation}
The space $\calK_{d_1,\ldots,d_n}$ corresponds to local degrees of 
freedom. They can be defined independently for each particle. So,\, 
$\calK_{d_1,\ldots,d_n}=\calK_{d_1}\otimes\cdots\otimes\calK_{d_n}$,\,
where $\calK_{d_m}$ is the space of ``subtypes'' (internal states) of the 
$m$-th particle. Like the type, the subtype of a particle is accessible by 
local measurements. However, it can be changed, while the type can not. 

The most interesting thing is the \emph{protected space} 
$\calM_{d_1,\ldots,d_n}$. It is not accessible by local measurements and 
is not sensitive to local perturbations, unless the particles come close to 
each other. This is an ideal place to store quantum information and operate 
with it. Unfortunately, the protected space does not have tensor product 
structure. However, it can be described as follows. Associated with each 
particle type $a$ is an irreducible representation $\calU_d$ of the quantum 
double $\calD$. Consider the product representation 
$\calU_{d_1}\otimes\cdots\otimes\calU_{d_n}$ and split it into components 
corresponding to different irreducible representations. The protected 
space is the component corresponding to the identity representation. 

If we swap two particles or move one around the other, the protected space 
undergoes some unitary transformation. Thus, the particles realize some 
multi-dimensional representation of the braid group. Such particles are 
called \emph{nonabelian anyons}. Note that braiding does not affect the local 
degrees of freedom. If two particles fuse, they can annihilate or 
become another particle. The protected space becomes smaller but some 
classical information comes out, namely, the type of the new particle. 
So, the we can do measurements on the protected space. Finally, if we 
create a new pair of particles of definite types, it always comes in a 
particular quantum state. So, we have a standard toolkit for quantum 
computation (new states, unitary transformations and measurements), except 
that the Hilbert space does not have tensor product structure. Universality 
of this toolkit is a separate problem, see Sec.~\ref{sec_univers}.

Our model gives rise to the same braiding and fusion rules as gauge field 
theory models~\cite{DPR,BDP}. The existence of local degrees of freedom 
(subtypes) is a new feature. These degrees of freedom appear because there is 
no explicit gauge symmetry in our model.

\section{Algebraic structure}

\subsection{Particles and local operators} \label{sec_locop}

This subsection is also rather abstract but the claims we do are concrete.
They will be proven in Sec.~\ref{sec_spaceL}.

As mentioned above, the ground state of the Hamiltonian~(\ref{Ham1}) is not
degenerate (on the sphere or on the plane regarded as an infinitely large
sphere). Excited states are characterized by their energies. The energy of an
eigenstate $|\psi\rangle$ is equal to the number of constraints
$(A(s)-1)|\psi\rangle=0$ or $(B(p)-1)|\psi\rangle=0$ which are
violated. Complete classification of excited states is a difficult
problem. Instead of that, we will try to classify \emph{elementary}
excitations, or particles.

Let us formulate the problem more precisely. Consider a few excited ``spots''
separated by large distances. Each spot is a small region where some of the
constrains are violated. The energy of a spot can be decreased by local
operators but, generally, the spot can not disappear.  Rather, it shrinks to
some minimal excitation (which need not be unique).  We will see (in
Sec.~\ref{sec_spaceL}) that any excited spot can be transformed into an
excitation which violates at most 2 constraints,\, $A(s)-1\equiv 0$ and
$B(p)-1\equiv 0$,\, where $s$ is an arbitrary vertex, and $p$ is an adjacent
face. Such excitations are be called
\emph{elementary excitations}, or \emph{particles}. Note that definition of
elementary excitations is a matter of choice. We could decide that an
elementary excitation violates 3 constraints. Even with our definition, the
``space of elementary excitations'' is redundant.

By the way, the space of elementary excitations is not well defined because
such an excitation does not exist alone. More exactly, the only one-particle
state on the sphere is the ground state. (This can be proven easily).  The
right thing is the space of two-particle excitations, $\calL(a,b)$. Here
$a=(s,p)$ and $b=(s',p')$ are the sites occupied by the particles. (Recall
that a site is a combination of a vertex and an adjacent face). The projector
onto $\calL(a,b)$ can be written as
$\prod_{r\not=s,s'}A(r)\,\prod_{l\not=p,p'}B(l)$.  Note that introducing a
third particle (say, $c$) will not give more freedom for any of the
two. Indeed, $b$ and $c$ can fuse without any effect on $a$.

Let us see how local operators act on the space $\calL(a,b)$. In this context,
a local operator is an operator which acts only on spins near $a$ (or near
$b$). Besides that, it should preserve the subspace $\calL(a,b)\subseteq\calN$
and its orthogonal complement. ($\calN$ is the space of all quantum states).
Example: the operators $A_g(a)$ and $B_h(a)$, where $a=(s,p)$, commute with
$A(r)$, $B(l)$ for all $r\not=s$ and $l\not=p$. Hence, they commute with the
projector onto the subspace $\calL(a,b)$. These operators generate an algebra
$\calD(a)\subset\LL(\calN)$. It will be shown in Sec.~\ref{sec_spaceL} that
$\calD(a)$ includes \emph{all} local operators acting on the space
$\calL(a,b)$,\, and the action of $\calD(a)$ on $\calL(a,b)$ is \emph{exact}
(i.e.\ different operators act differently).

Actually, the algebra $\calD(a)=\calD$ does not depend on $a$,\, only the
embedding $\calD\to\LL(\calN)$ does. This algebra is called the \emph{quantum
double} of the group $G$ and denoted by $\D(G)$. Its structure is determined
by the following relations between the operators $A_g=A_g(a)$ and $B_h=B_h(a)$
\begin{equation}
  A_f\,A_g \ =\  A_{fg} \qquad\qquad
  B_h\,B_i \ =\ \del_{h,i}\,B_h \qquad\qquad
  A_g\,B_h \ =\ B_{ghg^{-1}}A_g
\end{equation}
The operators $D_{(h,g)}=B_hA_g$ form a linear basis of $\calD$.  (In
\cite{DPR,BDP} these operators were denoted by\,
$\raisebox{0.5ex}{$\scriptstyle h$\,}
{\mathop{\mbox{\rule{0.4pt}{1.8ex}\rule{1.8ex}{0.4pt}}}\limits_g}$\,).  The
following multiplication rules hold
\[
   D_{(h_1,g_1)} D_{(h_2,g_2)} \ =\ 
  \del_{h_1,\,g_1h_2g_1^{-1}}\, D_{(h_1,\,g_1g_2)}
\]
This identity can be also written in a symbolic tensor form, with $h$ and $g$ 
being combined into one index:
\begin{equation} \label{locmult}
  \RD{m}\,\RD{n}\ =\ \OM{kmn}\, \RD{k}
  \qquad\qquad\quad
  \Om^{(h,g)}_{(h_1,g_1)\,(h_2,g_2)}\ =\,\ 
  \del_{h_1,\,g_1h_2g_1^{-1}}\ \del_{h,h_1}\ \del_{g,\,g_1g_2}
\end{equation}
(summation over {\boldmath$k$} is implied). Actually, $\calD$ is not only an
algebra, it is a quasi-triangular Hopf algebra, see 
Secs.~\ref{sec_ribop}, \ref{sec_further}.

Note that $\calD=\calD(a)$ is closed under Hermitian conjugation 
(in $\LL(\calN)$) which acts as follows
\begin{equation} \label{conjug}
  A_g^\dagger\,=\,A_{g^{-1}} \qquad\qquad   B_h^\dagger\,=\,B_h \qquad\qquad
  D_{(h,g)}^\dagger \,=\, D_{(g^{-1}hg,\,g^{-1})}
\end{equation}
Thus, $\calD=\calD(a)$ is a finite-dimensional $C^*$-algebra. Hence it has the
following structure
\begin{equation}
  \calD\ =\ \bigoplus_d \LL(\calK_d)
\end{equation}
where $d$ runs over all irreducible representations of $\calD$. We can 
interpret $d$ as particle's type.\footnote{
 \emph{Caution.} The local operators should not be interpreted as symmetry
 transformations. The true symmetry transformations, so-called
 \emph{topological operators}, will be defined in Sec.~\ref{sec_topop}.
 Mathematically, they are described by the same algebra $\calD$, but their
 action on physical states is different.}
The absence of particle corresponds to a certain one-dimensional
representation called the \emph{identity representation}. More exactly, the
operators $D_{(h,g)}$ act on the ground state $|\xi\rangle$ as follows
\begin{equation} \label{epsilon}
  \RD{k}\,|\xi\rangle\ =\ \EPS{k}\,|\xi\rangle \qquad\qquad
  \mbox{where}\quad \Eps_{(h,g)}\ =\ \del_{h,1}
\end{equation}

The ``space of subtypes'', $\calK_d$ actually characterize the redundancy of
our definition of elementary excitations. However, this redundancy is
necessary to have a nice theory of ribbon operators (see
Sec.~\ref{sec_ribop}).

Irreducible representations of $\calD$ can be described as follows~\cite{DPR}.
Let $u\in G$ be an arbitrary element, $C=\{gug^{-1}:\,g\in G\}$ its conjugacy
class,\, $E=\{g\in G:\,gu=ug\}$ its centralizer. There is one irreducible
representation $d=(C,\chi)$ for each conjugacy class $C$ and each irreducible
representation $\chi$ of the group $E$ (see below). It does not matter which
element $u\in C$ is used to define $E$. The conjugacy class $C$ can be
interpreted as magnetic charge whereas $\chi$ corresponds to electric
charge. For example, consider the group $S_3$ (the permutation group of order
3). It has 3 conjugacy classes of order 1, 2 and 3, respectively. So, the
algebra $\D(S_3)$ has irreducible representations of dimensionalities 1,1,2;
2,2,2; 3,3.

The simplest case is when $\chi$ is the identity representation, i.e.\ the
particle carries only magnetic charge but no electric charge. Then the
subtypes can be identified with the elements of $C$, i.e.\ the corresponding
space (denoted by $\calB_C$) has a basis $\{|v\rangle:\,v\in C\}$.  The local
operators act on this space as follows
\begin{equation} \label{vortrep}
  D_{(h,g)}|v\rangle\ =\ \del_{h,\,gvg^{-1}}\, |gvg^{-1}\rangle
\end{equation}

Now consider the general case. Denote by $W_f=W_f^{(\chi)}$ the irreducible
action of $f\in E$ on an appropriate space $\cal A_\chi$. Choose
arbitrary elements $q_v\in G$ such that $q_vuq_v^{-1}=v$ for each $v\in
C$. Then any element $g\in G$ can be uniquely represented in the form
$g=q_vf$, where $v\in C$ and $f\in E$. We can define a unique action of
$\calD$ on $\calB_C\otimes\calA_\chi$, such that
\begin{eqnarray}
  B_{h}\, \Bigl( |v\rangle\otimes|\eta\rangle \Bigr) &=& 
  \del_{h,v}\, |v\rangle\otimes|\eta\rangle        \\*
  A_{q_vf}\, \Bigl( |u\rangle\otimes|\eta\rangle \Bigr) &=& 
  |v\rangle\otimes\,W_f|\eta\rangle \qquad\quad (v\in C,\ f\in E) \nonumber
\end{eqnarray}
More generally,\, 
$D_{(h,g)}\Bigl(|v\rangle\otimes|\eta\rangle\Bigr)\,=\, 
 \del_{h,\,gvg^{-1}}\, |gvg^{-1}\rangle\otimes W_f|\eta\rangle$,\, 
where $f=q_v(q_{gvg^{-1}})^{-1}g$. This action is irreducible.

\subsection{Ribbon operators} \label{sec_ribop}

\begin{figure}[t]
\centerline{\epsfscale{0.7}\epsffile{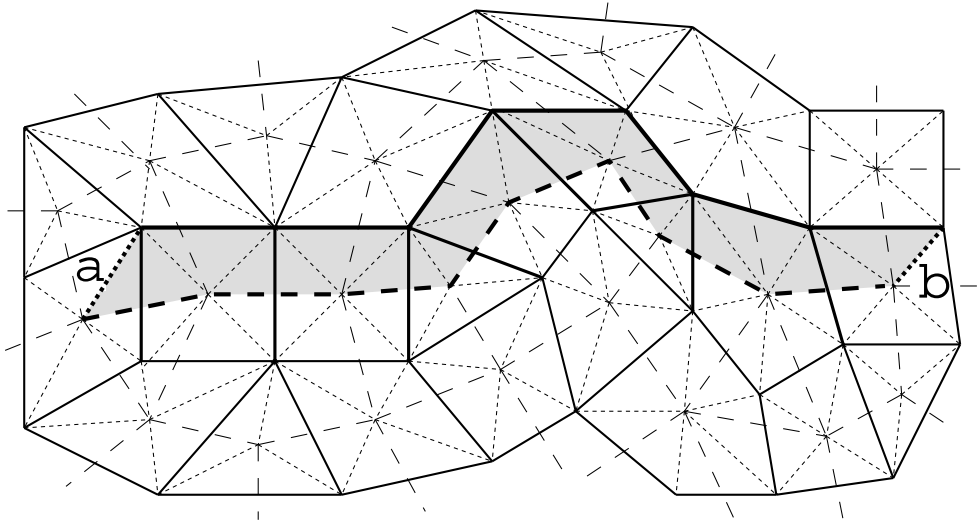}}
\caption{A ribbon on the lattice}
\label{fig_ribbon}
\end{figure}

The next task is to construct a set of operators which can create an arbitrary
two-particle state from the ground state. I do not know how to deduce an
expression for such operators; I will just give an answer and explain why it
is correct. In the abelian case (see Sec.~\ref{sec_abanyons}) there were two
types of such operators which corresponded to paths on the lattice and the
dual lattice, respectively.  In the nonabelian case, we have to consider both
types of paths together.  Thus, the operators creating a particle pair are
associated with a \emph{ribbon} (see fig.~\ref{fig_ribbon}). The ribbon
connects two sites at which the particles will appear (say, $a=(s,p)$ and
$b=(s',p')$). The corresponding operators act on the edges which constitute
one side of the ribbon (solid line), as well as the edges intersected by the
other side (dashed line).

For a given ribbon $t$, there are $N^2$ \emph{ribbon operators} $F^{(h,g)}(t)$
indexed by $g,h\in G$. They act as follows\footnote{
 Horizontal and vertical arrows are the two types of edges. Each of the two
 diagrams (6 arrows with labels) stand for a particular basis vector}
{
\unitlength=25mm
\def\relvshift{0.5}
\def\hsign#1{\vbox{\hbox to \unitlength {\hfil#1\hfil} \kern 2pt}}
\newcount\abc
\def\leftarrow(#1,#2)#3{\abc=#1 \advance\abc 1
     \put(\abc,#2){\vector(-1,0){1}} \put(#1,#2){\hsign{$#3$}}}
\def\uparrow(#1,#2)#3{\abc=#2 \advance\abc -1
     \put(#1,\abc){\vector(0,1){1}} \put(#1,\relvshift){\kern 2pt $#3$}}

\begin{equation} \label{ribop}
  \begin{array}{rl}
    F^{(h,g)}(t)\ & 
    \raisebox{-\relvshift\unitlength}{
      \begin{picture}(3,1.2)
        \leftarrow(0,1){x_1} \leftarrow(1,1){x_2} \leftarrow(2,1){x_3}
        \uparrow(0,1){y_1} \uparrow(1,1){y_2} \uparrow(2,1){y_3} 
      \end{picture}
    }
    \qquad = \bigskip \\
    =\quad \del_{g,\,x_1x_2x_3}\ &
    \raisebox{-\relvshift\unitlength}{
      \begin{picture}(3.3,1.2)
        \leftarrow(0,1){x_1} \leftarrow(1,1){x_2} \leftarrow(2,1){x_3}
        \uparrow(0,1){hy_1}
        \uparrow(1,1){x_1^{-1}hx_1\,y_2}
        \uparrow(2,1){(x_1x_2)^{-1}h(x_1x_2)\,y_3} 
      \end{picture}
    }
  \end{array}
\end{equation}
}

\noindent These operators commute with every projector $A(r)$, $B(l)$,
except for $r=s,s'$ and $l=p,p'$. This is the first important property of
ribbon operators. 

The operators $F^{(h,g)}(t)$ depend on the ribbon $t$. However, their action
on the space $\calL(a,b)$ depends only on the \emph{topological class} of the
ribbon This is also true for a multi-particle excitation space
$\calL(x_1,\ldots,x_n)$. More exactly, consider two ribbons, $t$ and $q$,
connecting the sites $x_1=a$ and $x_2=b$
\begin{figure}[h]
\centerline{\epsfscale{0.6}\epsffile{equiv.eps}}
\end{figure}
The actions of $F^{(h,g)}(t)$ and $F^{(h,g)}(q)$ on $\calL(x_1,\ldots,x_n)$
coincide provided none of the sites $x_3,\ldots,x_n$ lie on or between the
ribbons. This is the second important property of ribbon operators. We will
write $F^{(h,g)}(t)\equiv F^{(h,g)}(q)$, or, more exactly, 
$F^{(h,g)}(t)\stackrel{M}{\equiv}F^{(h,g)}(q)$, where $M=\{x_1,\ldots,x_n\}$.

Linear combination of the operators $F^{(h,g)}(t)$ are also called ribbon
operators. They form an algebra $\calF(t)\cong\calF$. The multiplication rules
are as follows
\begin{equation} \label{rib_mult}
  \RF{m}(t)\, \RF{n}(t)\ =\ \LAM{mnk}\, \RF{k}(t) \qquad\qquad
  \Lam^{(h_1,g_1)\,(h_2,g_2)}_{(h,g)}\ =\
  \del_{h_1h_2,\,h}\ \del_{g_1,g}\ \del_{g_2,g}
\end{equation}
(summation over {\boldmath $m$ and $n$} is implied).

Any ribbon operator on a long ribbon $t=t_1t_2$ (see figure below) can be 
represented in terms of ribbon operators corresponding to its parts, $t_1$ and
$t_2$
\begin{figure}[h]
\centerline{\epsfscale{0.6}\epsffile{comult.eps}}
\end{figure}
\begin{equation} \label{ribcomult}
  \RF{k}(t_1t_2)\ =\ \OM{kmn}\, \RF{m}(t_1)\, \RF{n}(t_2) \qquad\qquad
  \Om^{(h,g)}_{(h_1,g_1)\,(h_2,g_2)}\ =\ 
  \del_{g,\,g_1g_2}\ \del_{h_1,h}\ \del_{h_2,\,g_1^{-1}hg_1}
\end{equation}
(Note that $\RF{m}(t_1)$ and $\RF{n}(t_2)$ commute because the ribbons $t_1$
and $t_2$ do ton overlap). By some miracle, the tensor
$\Om^{\star}_{\star\star}$ is the same as in eg.~(\ref{locmult}). From the
mathematical point of view, eq.~(\ref{ribcomult}) defines a linear mapping\,
$\Delta(t_1,t_2):\ \calF(t_1,t_2)\to\calF(t_1)\otimes\calF(t_2)$,\,\, or just
$\Delta:\,\calF\to\calF$. Such a mapping is called a \emph{comultiplication}.

The comultiplication rules~(\ref{ribcomult}) allow to give another definition
of ribbon operators which is nicer than eq.~(\ref{ribop}). Note that a ribbon
consists of triangles of two types (see fig.~\ref{fig_ribbon}). Each triangle
corresponds to one edge. More exactly, a triangle with two dotted sides and
one dashed side corresponds to a combination of an edge and its endpoint, say,
$i$ and $r$. Similarly, a triangle with a solid side corresponds to a
combination of an edge and one of the adjacent faces, say, $j$ and
$l$. Each triangle can be considered as a short ribbon. The corresponding
ribbon operators are
\[
  F^{(h,g)}(i,r)\ =\ \del_{g,1}\,L^h(i,r) \qquad\qquad\quad
  F^{(h,g)}(j,l)\ =\ T^{g^{-1}}(j,l)
\]
The ribbon operators on a long ribbon can be constructed from these ones.

It has been already mentioned that the multiplication in $\calD$ and the
comultiplication in $\calF$ are defined by the same tensor
$\Om^{\star}_{\star\star}$. Actually, $\calD$ and $\calF$ are
Hopf algebras dual to each other. (For general account on Hopf algebras, see
\cite{Sw,Maj,Kas}).  The multiplication in $\calF$ corresponds to a 
comultiplication in $\calD$ defined as follows
\begin{equation} \label{loccomult}
  \Delta(\RD{k})\ =\ \LAM{mnk}\,\RD{m}\otimes\RD{n}
\end{equation}
(More explicitly,\,\, 
$\Delta(D_{(h,g)}) \,=\, \sum_{h_1h_2=h} D_{(h_1,g)}\otimes D_{(h_2,g)}\ $).
The unit element of $\calF$ is $1_\calF=\EPS{k}\RF{k}$, where 
$\EPS{k}$ are given by eq.~(\ref{epsilon});\, the tensor
$\Eps_{\star}$ also defines a counit of $\calD$ (i.e.\ the mapping
$\Eps:\calD\to\C$\,:\, $\Eps(\RD{k})=\EPS{k}$\,). The unit
of $\calD$ and the counit of $\calF$ are given by
\begin{equation}
  \rE^{(h,g)}\ =\ \del_{g,1}
\end{equation}
The Hopf algebra structure also includes an antipode, i.e.\ a mapping
$S:\calD\to\calD$\,:\, $S(\RD{k})=\RS{mk}\RD{m}$,\, or
$S:\calF\to\calF$\,:\, $S(\RF{m})=\RS{mk}\RF{k}$. The tensor 
$\rS^{\star}_{\star}$
is given by the equation
\begin{equation}
  \rS^{(h_1,g_1)}_{(h_2,g_2)}\ =\ 
  \del_{g_1^{-1}h_1g_1,\,h_2^{-1}}\ \del_{g_1,\,g_2^{-1}}
\end{equation}

Here is the complete list of Hopf algebra axioms.

\begin{equation} \label{axmult}
  \LAM{lmi}\ \LAM{ink}\ =\ \LAM{ljk}\ \LAM{mnj} \qquad\qquad\qquad
  \EPS{i}\ \LAM{imk}\ =\ \LAM{mjk}\ \EPS{j}\ =\ \DEL{mk}
\end{equation}

\begin{equation} \label{axcomult}
  \OM{ilm}\ \OM{kin}\ =\ \OM{klj}\ \OM{jmn} \qquad\qquad\qquad
  \RE{i}\ \OM{kin}\ =\ \OM{kmj}\ \RE{j}\ =\ \DEL{km}
\end{equation}

\begin{equation} \label{axconsist}
  \LAM{lmq}\ \OM{qkn}\ =\ \OM{lij}\ \OM{mrs}\ \LAM{irk}\ \LAM{jsn} \qquad\qquad
  \EPS{q}\ \OM{qkn}\ =\ \EPS{k}\,\EPS{n} \qquad\quad
  \LAM{lmq}\ \RE{q}\ =\ \RE{l}\,\RE{m}
\end{equation}

\begin{equation} \label{axanti}
  \RS{kl}\ \LAM{lmp}\ \OM{qkn}\ \DEL{nm}\ =\ 
  \DEL{kl}\ \LAM{lmp}\ \OM{qkn}\ \RS{nm}\ =\ \EPS{p}\,\RE{q}
\end{equation}
\medskip

Most of these identities correspond to physically obvious properties
of ribbon operators. Eq.~(\ref{axmult}) is a statement of the usual
multiplication axioms in the algebra $\calF$, namely,\,
$(\RF{l}\RF{m})\RF{n}=\RF{l}(\RF{m}\RF{n})$\, and\,
$1\RF{m}=\RF{m}1=\RF{m}$.  The first equation in~(\ref{axcomult})
(coassociativity of the comultiplication in $\calF$) can be proven by
expanding $\RF{k}(t_1t_2t_3)$ as\,
$\OM{kin}\,\RF{i}(t_1t_2)\,\RF{n}(t_3)$\, or\,
$\OM{klj}\,\RF{l}(t_1)\,\RF{j}(t_2t_3)$\, --- the result must be the
same.\footnote{ 
 The coassociativity is necessary and sufficient for that.  The
 sufficiency is rather obvious; the necessity follows from the fact
 that the mapping $\calF\to\calF(t)$ is injective, i.e.\ the operators
 $\RF{k}(t)$ with different {\boldmath$k$} are linearly independent.}
Eqs.~(\ref{axconsist}) mean that the multiplication and comultiplication
are consistent with each other. To prove the first equation
in~(\ref{axconsist}), expand\, $\RF{l}(t_1t_2)\,\RF{m}(t_1t_2)$\, in two
different ways. The second equation follows from the fact that
$\EPS{q}\,\RF{q}(t_1t_2)$ is the identity operator.

The antipode axiom~(\ref{axanti}) does not have explicit physical
meaning. Mathematically, it is a \emph{definition} of the tensor
$\rS^\star_\star$\,:\, the element 
$\Gam=\,\RS{lk}\,\RF{k}\otimes\RD{l}\,\in\calF\otimes\calD$\, is the inverse
to the canonical element $\Del=\RF{i}\otimes\RD{i}$.\, The antipode have the 
following properties which can be derived from~(\ref{axmult}--\ref{axanti})
\begin{equation} \label{propanti}
  \RS{il}\ \RS{jm}\ \LAM{lmp}\ =\ \LAM{jiq}\ \RS{qp} \qquad\qquad\quad
  \RS{pq}\ \OM{qij}\ =\ \OM{pml}\ \RS{mj}\ \RS{li}
\end{equation}

Finally, we can define a so-called \emph{skew antipode} $\rT^\star_\star$ as 
follows
\begin{equation} \label{skew}
  \RT{mi}\ \RS{in}\ =\ \RS{mj}\ \RT{jn}\ =\ \DEL{mn}
\end{equation}
In our case, $\RT{mi}=\RS{mi}$, but this is not true for a generic Hopf
algebra. The skew antipode have the following properties similar 
to~(\ref{axanti}) and~(\ref{propanti})
\begin{equation} \label{axskew}
  \RT{nl}\ \LAM{lmp}\ \OM{qkn}\ \DEL{km}\ =\ 
  \DEL{nl}\ \LAM{lmp}\ \OM{qkn}\ \RT{km}\ =\ \EPS{p}\,\RE{q}
\end{equation}
\begin{equation} \label{propskew}
  \RT{il}\ \RT{jm}\ \LAM{lmp}\ =\ \LAM{jiq}\ \RT{qp} \qquad\qquad\quad
  \RT{pq}\ \OM{qij}\ =\ \OM{pml}\ \RT{mj}\ \RT{li}
\end{equation}
(Note the distinction between~(\ref{axanti}) and~(\ref{axskew}), however).

The reader may be overwhelmed by a number of formal things, so let us
summarize what we know by now. We have defined two algebras, $\calD$ and
$\calF$, and their actions on the Hilbert space $\calN$. In this context, we
denote them by $\calD(a)$ and $\calF(t)$ because the actions depend on the
site $a$ or on the ribbon $t$, respectively. Operators from $\calD(a)$ affect
one particle whereas operators from $\calF(t)$ affect two particles. The
action of $\calF(t)$ on the space of $n$-particle states
$\calL(x_1,\ldots,x_n)$ depends only on the topological class of the ribbon
$t$. This space have not been found yet, even for $n=2$. (It will be found
after we learn more about local and ribbon operators). The algebra $\calF$ is
a Hopf algebra. The comultiplication allows to make up a long ribbon from
parts. There is a formal duality between $\calF$ and $\calD$. The
comultiplication in $\calF$ is dual to the multiplication in $\calD$. The
multiplication in $\calF$ is dual to a comultiplication in $\calD$. (The
meaning of the latter is not clear yet).

\subsection{Further properties of local and ribbon operators} 
\label{sec_further}

\begin{figure}[t]
\centerline{\epsfscale{0.6}\epsffile{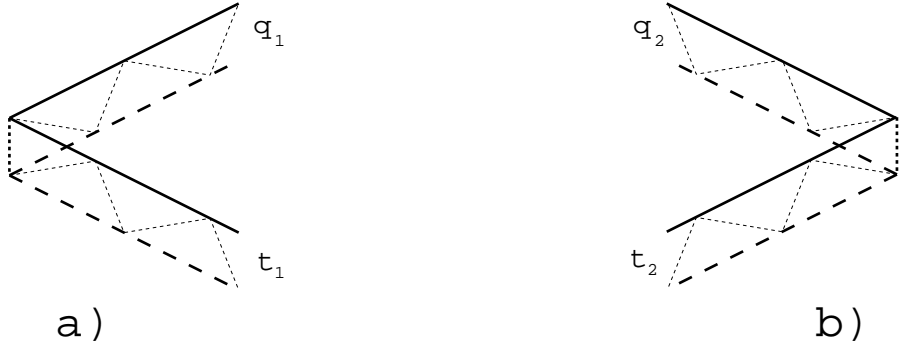}}
\caption{Two ribbons attached to the same site}
\label{fig_commute1}
\end{figure}

Let us study commutation relations between ribbon operators. Consider two
ribbons attached to the same site, as shown in fig.~\ref{fig_commute1}~a
or~b. Then
\[
  \begin{array}{rcl}
    F^{(h,g)}(t_1)\ F^{(v,u)}(q_1)\,&=&\, 
    F^{(v,u)}(q_1)\ F^{(v^{-1}hv,\,v^{-1}g)}(t_1)  \medskip\\
    F^{(h,g)}(t_2)\ F^{(v,u)}(q_2)\,&=&\,
    F^{(v,u)}(q_2)\ F^{(h,\,gu^{-1}vu)}(t_2)
  \end{array}
\]
In a tensor form, these equations read as follows
\begin{equation} \label{ribcommute}
  \begin{array}{rcl}
    \RF{m}(t_1)\ \RF{n}(q_1)\,&=&\, 
    \RR{ik}\ \OM{nij}\ \OM{mkl}\ \RF{j}(q_1)\ \RF{l}(t_1)  \bigskip\\
    \RF{m}(t_2)\ \RF{n}(q_2)\,&=&\,
    \RF{i}(q_2)\ \RF{k}(t_2)\ \OM{nij}\ \OM{mkl}\ \RI{jl}
  \end{array}
\end{equation}
where
\begin{equation} \label{R}
  \rR^{(h,g)(v,u)}\ =\ \del_{h,u}\ \del_{g,1} \qquad\qquad\quad
  \rI^{(h,g)(v,u)}\ =\ \del_{h^{-1}\!,\,u}\ \del_{g,1}  
\end{equation}
Note that 
\begin{equation} \label{Rinv}
  \RI{ik}\ \OM{nij}\ \OM{mkl}\ \RR{jl}\,\ =\,\ 
  \RR{ik}\ \OM{nij}\ \OM{mkl}\ \RI{jl}\,\ =\,\ \RE{n}\,\RE{m}
\end{equation}

\begin{figure}[t]
\centerline{\epsfscale{0.6}\epsffile{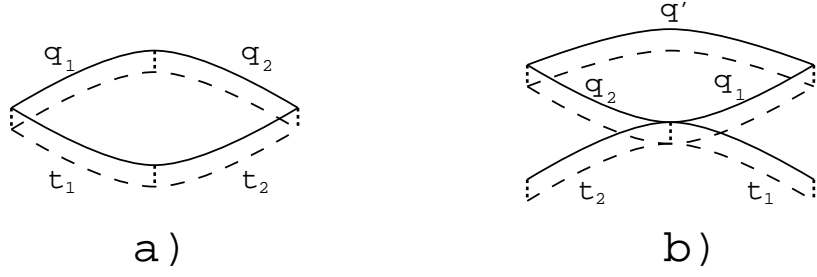}}
\caption{Checking consistency of the commutation relations}
\label{fig_commute2}
\end{figure}

To prove\footnote{ 
 This proof is not rigorous, but an interested reader can
 easily fix it. Anyway, you can just substitute (\ref{R}) into (\ref{Rinv})
 and check it directly.}
(and to see the physical meaning of) this equation, consider the configuration
shown in fig.~\ref{fig_commute2}b. Clearly, $\RF{r}(t_2t_1)$ and $\RF{s}(q')$
commute. On the other hand, $\RF{s}(q')\equiv\RF{s}(q_2q_1)$, so
$\RF{r}(t_2t_1)$ and $\RF{s}(q_2q_1)$ commute. It follows that\,
$\RI{ik}\,\OM{nij}\,\OM{mkl}\,\RR{jl}\,=\,\RE{n}\RE{m}$.\, This identity can
be easily written in an invariant form, namely, $\bar
RR=1_{\calD\otimes\calD}$,\, where $R=\RR{jl}\,\RD{j}\otimes\RD{l}$\, and
$\bar R=\RI{ik}\,\RD{i}\otimes\RD{k}$. It also implies that $R\bar
R=1_{\calD\otimes\calD}$ because the algebra $\calD\otimes\calD$ is finite
dimensional. Thus, $\bar R=R^{-1}$.

The tensor $\rR^\star_\star$ (or the element $R\in\calD\otimes\calD$) is
called the \emph{$R$-matrix}. It satisfies the following axioms
\begin{equation} \label{axR1}
  \LAM{ijk}\ \RR{km}\ =\ \RR{il}\ \RR{jn}\ \OM{mln} \qquad\qquad\qquad
  \RR{mk}\LAM{jik}\ =\ \OM{mln}\ \RR{li}\ \RR{nj}
\end{equation}
\begin{equation} \label{axR2}
  \LAM{jik}\ =\ \OM{i{lm}r}\ \OM{j{pn}s}\ \RR{lp}\ \LAM{mnk}\ \RI{rs}
\end{equation}
where $\OM{i{lm}r}=\OM{ilu}\OM{umr}=\OM{ulm}\OM{iur}$. Eqs.~(\ref{axR1})
follow from~(\ref{ribcommute}). Conversely, these equations ensure that the
commutation relation are consistent. To prove the first equation
in~(\ref{axR1}), commute $\RF{m}(t_1)\,\RF{i}(q_1)\,\RF{j}(q_1)$ in two
different ways. You will get
$\mathrm{W}^\boldind{ijm}_\boldind{ab}\,\RF{a}(q_1)\,\RF{b}(t_1)$, with two
different expressions for $\mathrm{W}^\boldind{ijm}_\boldind{ab}$. Then
calculate $\mathrm{W}^\boldind{ijm}_\boldind{ab}\,\RE{a}\,\RE{b}$\, using the
axioms~(\ref{axmult}--\ref{axconsist}). The second equation in~(\ref{axR1})
can be proven in a similar way.

To prove eq.~(\ref{axR2}), consider the configuration shown in
fig.~\ref{fig_commute2}a. Clearly, $\RF{i}(q_1q_2)\equiv\RF{i}(t_1t_2)$,\,
so\, $\RF{j}(t_1t_2)\,\RF{i}(q_1q_2)\,\equiv\,\LAM{jik}\,\RF{k}(t_1t_2)$.\, 
On the other hand, we can first expand $\RF{j}(t_1t_2)$ and $\RF{i}(q_1q_2)$
using the comultiplication rules, and then apply the commutation
relations~(\ref{ribcommute}). The result must be the same.

Let $t$ be a ribbon connecting sites $a$ and $b$. The local and ribbon 
operators commute as follows
\begin{figure}[h]
\centerline{\epsfscale{0.6}\epsffile{commute3.eps}}
\end{figure}
\begin{equation} \label{commrel}
  \RF{m}(t)\ \RD{i}(a)\ =\ \LAM{jki}\ \OM{mkl}\ \RD{j}(a)\ \RF{l}(t)
  \qquad\qquad
  \RD{i}(b)\ \RF{m}(t)\ =\ \OM{mlk}\ \LAM{kji}\ \RF{l}(t)\ \RD{j}(b)
\end{equation}
These commutation relations can be also written in the form
\begin{equation}
  \begin{array}{rcl}
    \RD{j}(a)\ \RF{l}(t) &\,=\,&
    \LAM{ikj}\ \RT{nk}\ \OM{lnm}\ \RF{m}(t)\ \RD{i}(a) \medskip\\
    \RF{l}(t)\ \RD{j}(b) &\,=\,&
    \OM{lmn}\ \RT{nk}\ \LAM{kij}\ \RD{i}(b)\ \RF{m}(t)
  \end{array}
\end{equation}
where $\rT^\star_\star$ is the skew antipode 
(see eqs.~(\ref{skew},\ref{axskew})).

Finally, we introduce some special elements $C\in\calD$ and $\Tau\in\calF$.
The first one has a clear physical meaning:\, the corresponding operator
$C(a)=A(a)\,B(a)$ projects out states with no particle at the site $a$. The
element $C$ can be represented in the form
\begin{equation} \label{C}
  C\ =\ \RC{i}\,\RD{i} \qquad\qquad
  \mbox{where}\quad \rC^{(h,g)}\,=\,N^{-1}\del_{h,1}
\end{equation}
It has the following properties:
\[
  CX\, =\ XC\, =\, \Eps(X)\,C \quad \mbox{for any}\,\ X\in\calD\,
 \qquad\qquad\qquad  \Eps(C)\, =\, 1
\]
or, in tensor notations,
\begin{equation}
  \OM{kij}\,\RC{i}\ =\ \OM{kji}\,\RC{i}\ =\ \EPS{j}\,\RC{k} 
  \qquad\qquad\qquad  \EPS{k}\,\RC{k}\ =\ 1
\end{equation}

The element $\Tau\in\calF$ is dual to $C$; it is defined as follows
\begin{equation} \label{Tau}
  \Tau\ =\ \TAU{i}\,\RF{i} \qquad\qquad
  \mbox{where}\quad \Tau_{(h,g)}\,=\,N^{-1}\del_{1,g}
\end{equation}
Its properties are as follows
\begin{equation}
  \LAM{ijk}\,\TAU{i}\ =\ \LAM{jik}\,\TAU{i}\ =\ \RE{j}\,\TAU{k}
  \qquad\qquad\qquad \RE{k}\,\TAU{k}\ =\ 1
\end{equation}
Note that $\TAU{k}\,\RC{k}=N^{-2}$. Using these properties, we can we can
derive an important consequence from the commutation relations~(\ref{commrel})
\begin{equation} \label{relC}
  \TAU{s}\,\OM{smp}\,\RS{pq}\ \RF{m}(t)\, C(a)\, \RF{q}(t)\,\ =\,\ 
  \TAU{s}\,\OM{spm}\,\RS{pq}\ \RF{q}(t)\, C(b)\, \RF{m}(t)\,\ =\,\  N^{-2}
\end{equation}

\subsection{The space $\calL(a,b)$} \label{sec_spaceL}

Now we are in a position to find the space $\calL(a,b)$ and to prove the
assertions from Sec.~\ref{sec_locop}. The first assertion was that any excited
spot can be transformed into one particle. It is simple if we can transform 
two particles into one by ribbon operators. Let us choose an arbitrary site 
$b$ the excited spot to be compressed to. Let some constraint, 
$A(s)-1\equiv 0$ or $B(p)-1\equiv 0$, be violated. Choose any site $a$ 
containing the vertex $s$ or the face $p$. Connect $a$ and $b$ by a ribbon. 
By the assumption, we can clean up the site $a$ while changing the state of
$b$, but without violating any more constraint. We can repeat this procedure
again and again to clean up the whole spot.

So, we only need to show that two particles can be transformed into one.  What
does it mean exactly? Physically, any transformation must be unitary, but it
can involve also some external system. (Otherwise, it is impossible to
``decrease entropy'', i.e.\ to convert many states into fewer).  On the other
hand, it is clear that unitarity is not relevant to this problem.  However, we
should exclude degenerate transformations, such as multiplication by the zero
operator. So, it is better to reformulate the assertion as follows:\, any
two-particle state (plus some other excitations far away) can be obtained from
one-particle states (plus the same excitations far away). Let
$|\psi\rangle\in\calL(a,b,\ldots)$ be such a two-particle state. We are going
to use the formula~(\ref{relC}). Let
\begin{equation} \label{decomp}
  G_\boldind{q}\ =\ N^2\,\TAU{s}\,\OM{smp}\,\RS{pq}\ \RF{m}(t)
  \qquad\qquad\quad
  |\eta^\boldind{q}\rangle\ =\ C(a)\, \RF{q}(t)\, |\psi\rangle
\end{equation}
Then\, $|\psi\rangle\,=\,G_\boldind{q}\,|\eta^\boldind{q}\rangle$. The states
$|\eta^\boldind{q}\rangle$ belong to $\calL(b,\ldots)$, i.e.\ do not contain
excitation at $a$. This is exactly what we need.

The other two assertions were about the action of local operators on the space
$\calL(a,b)$, so we need to find this space first. We can consider this space 
as a representation of the algebra $\calE\cong\calE(t)$ generated by the
operators\, $\RD{j}=\RD{j}(a)$,\,\, $\RF{l}=\RF{l}(t)$\, and 
$\RD{j}'=\RD{j}(b)$. As a linear space,
$\calE=\calD\otimes\calF\otimes\calD$. (Thus, $\calE$ has dimensionality
$N^6$). Multiplication in $\calE$ is defined by the commutation
relations~(\ref{commrel}). We will call $\calE\cong\calE(t)$ the algebra of
\emph{extended ribbon operators}. It is just an algebra, not a Hopf algebra. 
More exactly, it is a finite-dimensional $C^*$-algebra. The involution 
($=$Hermitian conjugation) is given by the formulas (cf.~(\ref{conjug}))
\begin{equation} \label{conjug1}
  (D_{(h,g)})^\dagger\ =\ D_{(g^{-1}hg,\,g^{-1})} \qquad\quad
  (F^{(h,g)})^\dagger\ =\ F^{(h^{-1},g)} \qquad\quad
  (D'_{(h,g)})^\dagger\ =\ D'_{(g^{-1}hg,\,g^{-1})}
\end{equation}

\noindent[\emph{Remark.}\, Apparently, the algebra $\calE$ will play the
central role in a general theory of topological quantum order. Indeed, we were
lucky to define ribbon operators separately from local operators. In the
general case, ribbon operators should be mixed with local operators.]

So, we are looking for a particular representation $\calL$ of the algebra
$\calE$. This representation must contain a special vector $|\xi\rangle$ (the
ground state) such that
\begin{equation}
  \RD{k}\,|\xi\rangle\ =\ \EPS{k}\,|\xi\rangle \qquad\qquad\quad
  \RD{k}'\,|\xi\rangle\ =\ \EPS{k}\,|\xi\rangle
\end{equation}
We start with constructing a representation $\check\calL$ spanned by the
vectors $\PSI{k}\,=\,\RF{k}\,|\xi\rangle$. (It will be proven after that
$\check\calL=\calL$). We assume that the vectors $\PSI{k}$ are linearly
independent. This need not be the case in the representation $\calL$ but we
can postulate $\PSI{k}$ being linearly independent in $\check\calL$. Thus,
$\calL$ contains a factor-representation of $\check\calL$.

The representation $\check\calL$ is given by the formulas
\begin{equation} \label{repres1}
  \RD{j}\,\PSI{k}\ =\ \RT{nj}\,\OM{knm}\,\PSI{m} \qquad\quad
  \RF{j}\,\PSI{k}\ =\ \LAM{jkm}\,\PSI{m} \qquad\quad
  \RD{j}'\,\PSI{k}\ =\ \OM{kmj}\,\PSI{m}
\end{equation}
It is easy to show that this representation is irreducible. Hence, $\calL$
contains $\check\calL$, i.e.\ the vectors $\PSI{k}$ are linearly independent
in $\calL$. The scalar products between the vectors $\PSI{k}$ can be found
from~(\ref{repres1}) and~(\ref{conjug1}),
\begin{equation}
  \langle\psi^{(v,u)}|\psi^{(h,g)}\rangle\ =\ N^{-1}\,\del_{v,h}\,\del_{u,g}
\end{equation}

To prove that $\check\calL=\calL$, we use the equation~(\ref{relC}) again.
For an arbitrary two-particle state $|\psi\rangle\in\calL$, define the vectors
$|\eta^\boldind{q}\rangle$ and operators $G_\boldind{q}$ as in
eq.~(\ref{decomp}). Then
$|\psi\rangle\,=\,G_\boldind{q}\,|\eta^\boldind{q}\rangle$. One could say that
$|\eta^\boldind{q}\rangle\in\calL(b)$ but, actually, the space $\calL(b)$ is
spanned by the sole vector $|\xi\rangle$. It follows that
$|\psi\rangle\in\check\calL$\, --- the assertion has been proven. Thus, the
action of local and ribbon operators on the space $\calL=\calL(a,b)$ is given
by eq.~(\ref{repres1}).

It is easy to see that the action of $\calD(a)$ on $\calL(a,b)$ is exact
(though it is reducible). Besides that, $\calD(a)$ is the commutant of
$\calD(b)$ in $\LL(\calL(a,b))$ and \emph{vise versa}. (That is, $\calD(a)$
consists of all operators $X\in\LL(\calL(a,b))$ which commute with every
$Y\in\calD(b)\,$). Hence, $\calD(a)$ includes {\em all} local operators acting
on the space $\calL(a,b)$. Indeed, a local operator, which involves only spins
near the site $a$, must commute with any operator acting on distant spins. Of
course, there are many such operators, but their action on the two-particle
space $\calL(a,b)$ coincides with the action of the operators from $\calD(a)$.
This is also true for a multi-particle excitation space
$\calL(x_1,\ldots,x_n)$.

The space $\calL(x_1,\ldots,x_n)$ can be described as follows. Let us connect
the sites $x_1,\ldots,x_n$ by $n-1$ ribbons $t_1,\ldots,t_{n-1}$ in an
arbitrary way so that the ribbons form a tree.  Then the vectors 
$\PSI{k_1,\ldots,k_{n-1}}\,=\,
 \RF{k_1}(t_1)\ldots\RF{k_{n-1}}(t_{n-1})\,|\xi\rangle$
form a basis of $\calL(x_1,\ldots,x_n)$. Choosing different ribbons means
choosing a different basis. In the next section we will give another
description of multi-particle excitation spaces.

\section{Topological operators, braiding and fusion} \label{sec_topop}

Let us consider again the $n$-particle excitation space 
$\calL=\calL(x_1,\ldots,x_n)$. The algebra $\LL(\calL)$ includes the local
operator algebras $\calD(x_1),\ldots,\calD(x_1)$. An operator $Y\in\LL(\calL)$
which commute with every $X\in\calD(x_j)$\,\, ($j=1,\ldots,n$)\,\, is called a
\emph{topological operator}. Physically, topological operators correspond to
nonlocal degrees of freedom. For $n=2$, the algebra of topological operators
coincides with the center of $\calD(x_1)$ or $\calD(x_2)$. (The two centers
coincide). Hence, the only nonlocal degree of freedom is the type of either
particle. (The two particles correspond to dual representations of $\calD$; in
other words, these are a particle and an anti-particle). So, there is no hidden
(i.e.\ quantum nonlocal) degree of freedom in this case. Such hidden degrees
of freedom appear for $n\ge 3$.

\begin{figure}[t]
\centerline{\epsfscale{0.6}\epsffile{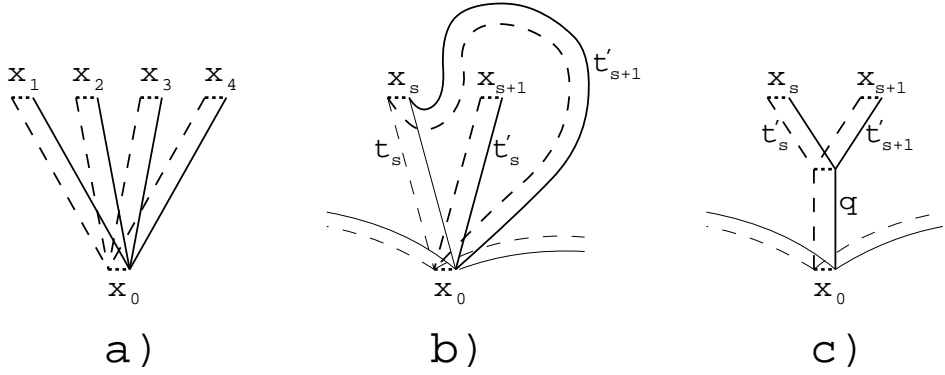}}
\caption{Braiding and fusion in terms of ribbon transformations}
\label{fig_braidfus}
\end{figure}

To describe the space $\calL$ and operators acting on it, let us choose an
arbitrary site $x_0$ (distinct from $x_1,\ldots,x_n$) and connect it with
$x_1,\ldots,x_n$ by non-intersecting ribbons $t_1,\ldots,t_n$, see
fig.~\ref{fig_braidfus}a.  As stated above, the space
$\calL(x_0,x_1,\ldots,x_n)$ is spanned by the vectors
\begin{equation} \label{ribbasis}
  \PSI{k_1,\ldots,k_{n}}\ =\ \RF{k_1}(t_1)\ldots\RF{k_{n}}(t_n)\,|\xi\rangle
\end{equation}
The space in question, $\calL=\calL(x_1,\ldots,x_n)$ is contained in
$\calL(x_0,x_1,\ldots,x_n)$. It consists of all vectors
$|\psi\rangle\in\calL(x_0,x_1,\ldots,x_n)$ which are invariant under the
action of $\calD(x_0)$ on the latter space.

The advantage of this description is that we can easily find all operators on
the space $\calL(x_0,x_1,\ldots,x_n)$ which commute with
$\calD(x_1)\otimes\ldots\otimes\calD(x_n)$. These are simply operators which
act on the ends of the ribbons $t_1,\ldots,t_n$ attached to the site $x_0$.
More exactly, an operator $\RD{j}^{(r)}$\,\, $(r=1,\ldots,n)$\, acts on the
$r$-th ribbon as $\RD{j}'=\RD{j}(x_0)$ (see~eq.(\ref{repres1})), but does not
affect the other ribbons,
\begin{equation} \label{topop}
  \Bigl(\RD{j_1}^{(1)}\otimes\ldots\otimes\RD{j_n}^{(n)}\Bigr)\ 
  \PSI{k_1,\ldots,k_{n}}\,\ =\,\ 
  \OM{{k_1}{m_1}{j_1}}\ldots\OM{{k_n}{m_n}{j_n}}\,\PSI{m_1,\ldots,m_{n}}
\end{equation}
Thus we arrive to an interesting physical conclusion. Let us consider only one
particle attached to an end of a semi-infinite ribbon (an analog of Dirac's
string). Then \emph{the topological operators act on the far end of the
ribbon}.
\smallskip

\noindent\textbf{Example.}\, Let us see how the topological operators act on
magnetic vortices. As shown in Sec.~\ref{sec_locop}, a vortex type is
characterized by a conjugacy class $C$ of the group $G$. Individual
topological states of the particle are characterized by particular elements
$v\in C$. In terms of the notation~(\ref{ribbasis}), such a state can be
represented as follows
\[
  |u,v\rangle\ =\ |C|^{1/2} \sum_{x:\ x^{-1}ux=v}\, |\psi^{(u,x)}\rangle
\]
where $u\in C$ characterize the local state of the particle. One can easily
check that $D_{(h,g)}'|u,v\rangle\,=\,\delta_{h,\,gvg^{-1}}|u,h\rangle$. This
is consistent with eq.~(\ref{vortrep}). Note that the local degree of freedom,
$u$, is not affected.

How can we physically apply topological operators to particles? We can just
move the particles around each other; this process is called \emph{braiding}.
Let us see what happens if we interchange two particles, $x_s$ and $x_{s+1}$,
counterclockwise, as shown in fig.~\ref{fig_braidfus}b. The state
$\PSI{\ldots,k,l,\ldots}$ becomes a new state
\[
  \PSI{\ldots,k,l,\ldots}_{\raisebox{-0.3ex}{\scriptsize new}}
  \,\ =\,\  \calR_\curvearrowleft\, \PSI{\ldots,k,l,\ldots}\,\ =\,\ 
  \ldots\RF{k}(t'_s)\ \RF{l}(t'_{s+1})\ldots\,|\xi\rangle
\]
To represent this state in the old basis, we should represent the operator
$\RF{k}(t'_s)\,\RF{l}(t'_{s+1})$ in terms of $\RF{m}(t_s)$ and
$\RF{n}(t_{s+1})$. Obviously, $\RF{k}(t'_s)=\RF{k}(t_{s+1})$;\, also
$\RF{l}(t'_{s+1})\equiv\RF{l}(t_s)$ as long as there is no particle
at $x_{s+1}$, i.e.\ the operator $\RF{k}(t_{s+1})$ is not applied yet. Hence
\[
  \RF{k}(t'_s)\ \RF{l}(t'_{s+1})\ \equiv\ 
  \RF{k}(t_{s+1})\ \RF{l}(t_s)
\]
Now we can apply the second commutation relation from~(\ref{ribcommute}). 
(Actually, we should reverse it). It follows that
\[
  \RF{k}(t_{s+1})\ \RF{l}(t_s)\,\ =\,\ 
  \RR{ji}\ \OM{lmi}\ \OM{knj}\
  \RF{m}(t_{s})\ \RF{n}(t_{s+1})  
\]
\[
  \calR_\curvearrowleft\, \PSI{\ldots,k,l,\ldots}
  \,\ =\,\
  \RR{ji}\ \Bigl(\RD{i}^{(s)}\otimes\RD{j}^{(s+1)}\Bigr)\ 
  \PSI{\ldots,l,k,\ldots}
\]
(see eq.~(\ref{topop})). Consequently, the counterclockwise interchange
operator has the form
\begin{equation} \label{interchange}
  \calR_\curvearrowleft\ =\ \RR{ji}\ (\RD{i}'\otimes\RD{j}')\ \Sig\ =\  
  \Sig\ \RR{ij}\, (\RD{i}'\otimes\RD{j}')
\end{equation}
where $\Sig$ is the permutation operator, and $\RD{i}'$, $\RD{j}'$ are
understood as topological operators. (Note that the operator $\Sig$ permutes
both topological and local degrees of freedom).
\smallskip

\noindent\textbf{Example.}\, Consider two magnetic vortices characterized by
topological parameters $v_1,v_2\in G$. The operator $\calR\curvearrowleft$ 
acts on the state $|v_1,v_2\rangle$ as follows
\begin{equation} \label{vortinterchange}
  \calR_\curvearrowleft\,|v_1,v_2\rangle\ =\ |v_1v_2v_1^{-1},\,v_1\rangle
\end{equation}
(The local parameters, $u_1$ and $u_2$, are suppressed in this formula).
\smallskip

Finally, let us see what happens if two particles, $x_s$ and $x_{s+1}$, fuse
into one. The resulting particle can be characterized by the action of
topological operators on it. From this point of view, we can just glue parts
of the corresponding ribbons (see fig.~\ref{fig_braidfus}c) instead of fusing
the particles themselves. Then
\begin{eqnarray}
  \RF{k}(t_s)\ \RF{l}(t_{s+1}) &\equiv&
  \OM{kmi}\ \OM{lnj}\ \LAM{ijp}\ \RF{m}(t'_s)\ \RF{n}(t'_{s+1})\ \RF{p}(q)
\nonumber\\
  |\psi^{\ldots,k,l,\ldots}\rangle &\equiv&
  \OM{kmi}\ \OM{lnj}\ \LAM{ijp}\ \RF{m}(t'_s)\ \RF{n}(t'_{s+1})\ 
  |\psi^{\ldots,p,\ldots}\rangle
\nonumber\\
  \LAM{uvr}\ \RD{u}^{(s)}\otimes\RD{v}^{(s+1)} &\equiv&
  \RD{r}'
\nonumber
\end{eqnarray}
where $\RD{r}'$ acts on the end of the ribbon $q$. Thus, fusion is described
by the comultiplication in the algebra $\calD$, see equation~(\ref{loccomult}).
(To avoid confusion, one should replace $D_\star$ with $D_\star'$ in that
equation).  The topological operator $\Delta(\RD{k}')$ acts on a particle pair
as the topological operator $\RD{k}'$ on the particle resulting from fusion.
\smallskip

\noindent\textbf{Example.} Consider a pair of opposite magnetic vortices 
$|v,\,v^{-1}\rangle$. The operators $\Delta(\RD{k}')$ act on this state as
follows
\begin{equation} \label{vortfusion}
  \Delta(D_{(h,g)}')\ |v,\,v^{-1}\rangle\ =\ 
  \del_{h,1}\ |gvg^{-1},\,gv^{-1}g^{-1}\rangle
\end{equation}
It terms of the representation classification (see Sec.~\ref{sec_locop}), this
action corresponds to the pair $(C,\chi)$, where $C=\{1\}$, and $\chi$ is the
adjoint representation of $G$. Thus, when opposite magnetic vortices fuse, the
resulting particle has no magnetic charge but may have some electric charge.

\section{Universal computation by anyons} \label{sec_univers}
(This section should be considered as an abstract of results to be presented
elsewhere).
\bigskip

Universal quantum computation is possible in the model based on the
permutation group $S_5$. (Unsolvability of the group seems to be important).
Vortex pairs $|v,\,v^{-1}\rangle$, where $v$ is a transposition, are
used as qubits. It is possible to perform the following operations.
\begin{enumerate}
\item 
  To produce pairs with zero charge. If a pair is created from the ground
  state, it has no charge automatically.
\item 
  To measure the electric charge of a vortex pair destructively. For this, we
  should simply fuse the the pair into one particle.
\item \label{classtrans}
  To perform the following unitary transformation on two pairs
\begin{equation}
  |u,\,u^{-1}\rangle\, \otimes\, |v,\,v^{-1}\rangle\,\ \mapsto\,\ 
  |vuv^{-1},\,vu^{-1}v^{-1}\rangle\, \otimes\, |v,\,v^{-1}\rangle
\end{equation}
  For this, we pull the first pair (as a whole) between the particles of the
  second pair.
\item \label{classmeas}
  To measure the value of $v$ and produce an unlimited number of pure states
  $|v,\,v^{-1}\rangle$ for any given transposition $v$ (say, $(1,2)$ or
  $(2,3)\,$). [At first sight, it is impossible because we can only measure
  the conjugacy class of a $v$. However, we can agree on a given state to
  correspond to $v=(1,2)$. Then we use it as a reference to produce an
  unlimited number of copies.]
\end{enumerate}

The operations~\ref{classtrans} and~\ref{classmeas} are sufficient to perform
universal classical computation. It is relatively simple to run quantum
algorithms based on measurements~\cite{Kit3}. Simulating a universal gate set
is more subtle and requires \emph{composite qubits}. That is, a usual qubit
(with two distinct states) is represented by several vortex pairs.

\section*{Concluding remarks}

It has been shown that anyons can arise from a Hamiltonian with local
interactions but without any symmetry. These anyons can be used to perform
universal quantum computation. There are still many things to do and questions
to answer. First of all, it is desirable to find other models with anyons
which allow universal quantum computation. (The group $S_5$ is quite
unrealistic for physical implementation). Such models must be based on a more
general algebraic structure rather than the quantum double of a group
algebra. A general theory of anyons and topological quantum order is lacking.
[In a sense, a general theory of anyons already exists~\cite{DPR}; it is based
on quasi-triangular quasi-Hopf algebras. However, this theory either merely
postulates the properties of anyons or connects them to certain field
theories. This is quite unlike the theory of local and ribbon operators which
describes both the properties of excitations and the underlying spin
entanglement.] It is also desirable to formulate and prove some theorem about
existence and the number of local degrees of freedom. (It seems that the local
degrees of freedom are a sign that anyons arise from a system with no symmetry
in the Hamiltonian). Finally, general understanding of dynamically created, or
``materialized'' symmetry is lacking. There one may find some insights for
high energy physics. If we adopt a conjecture that the fundamental Hamiltonian
or Lagrangian is not symmetric, we can probably infer some consequences about
the particle spectrum.
\bigskip

\noindent\textbf{\large Acknowledgements.} I am grateful to J.~Preskill,
D.~P.~DiVincenzo and C.~H.~Bennett for interesting discussions and questions
which helped me to clarify some points in my constructions. This work was
supported, in part, by the Russian Foundation for Fundamental Research, grant
No~96-01-01113. Part of this work was completed during the 1997 
Elsag-Bailey -- I.S.I. Foundation research meeting on quantum computation.

\end{document}